\documentclass[aps,prd,amsmath,amssymb,superscriptaddress,altaffillsymbol, preprintnumbers,preprint,nofootinbib,a4paper]{revtex4}
\pdfoutput=1
\usepackage{amsthm}
\usepackage{amsmath}
\usepackage{graphicx}
\usepackage{color}
\usepackage{subfigure}
\usepackage{multirow}

\usepackage{dcolumn}
\usepackage{bm}
\usepackage{slashed} 
\usepackage{epsfig}
\usepackage{hyperref}
\usepackage{verbatim}

\usepackage{mathtools}
\usepackage{xspace}

\usepackage[pdf]{pstricks}
\usepackage{epstopdf}

\newcommand{\beq}{\begin{eqnarray}}
\newcommand{\eeq}{\end{eqnarray}}

\newcommand{\bmp}{\noindent\begin{minipage}{16cm}}
\newcommand{\emp}{\end{minipage}\vskip 7mm} 

\newcommand{\GeV}{\mbox{ ${\mathrm{GeV}}$}}
\newcommand{\TeV}{\mbox{ ${\mathrm{TeV}}$}}

\newcommand{\uni}{\mbox{ ${\bf 1}_{2\times 2}$ }}
\newcommand{\Tr}{\mbox{Tr}\;}
\newcommand{\be}{\begin{eqnarray}}
\newcommand{\ee}{\end{eqnarray}}

\newcommand{\st}{s_\theta}
\newcommand{\ct}{c_\theta}

\newcommand{\ctt}{c_{2\theta}}
\newcommand{\SU}{\mbox{SU}}

\newcommand{\ha}{\hat{a}}
\newcommand{\id}{1}

\newcommand{\ii}{\mathrm{i}}
\newcommand{\ord}{\mathcal{O}}
\newcommand{\mO}{\mathcal{O}}
\newcommand{\mL}{\mathcal{L}}
\newcommand{\mM}{\mathcal{M}}

\newcommand{\ez}{\Sigma_0} 
\newcommand{\hc}{\text{h.c.}}

\def\eq#1{{eq.~(\ref{#1})}}
\def\eqs#1#2{{eqs.~(\ref{#1})--(\ref{#2})}}
\def\fig#1{{fig.~(\ref{#1})}}
\def\sec#1{{sec.~(\ref{#1})}}
\def\tab#1{{tab.~(\ref{#1})}}

\def\app#1{{appendix.~(\ref{#1})}}

\newcommand{\mg}{\textsc{MG5\_aMC@NLO}\xspace}
\newcommand{\feynrules}{\textsc{FeynRules}\xspace}

\begin{document}
\title{Towards the precise description of Composite Higgs models at colliders}

\author{Diogo Buarque Franzosi}
\affiliation{Stockholm University, Department of Physics,
106 91 Stockholm, Sweden}
\email{diogo.buarque.franzosi@gmail.com}
 

\begin{abstract}

We present a framework to study the interactions among Nambu-Goldstone bosons (NGB), pseudo-NGB (pNGBs) and gauge bosons in Composite Higgs (CH) models at high energies,  
including operators of order $\mathcal{O}(p^4)$ and $\mathcal{O}(p^2g^2)$ in the chiral expansion and topological terms. 
The set of (p)NGBs comprises the longitudinal modes of electroweak bosons, the Higgs boson, and possibly other scalar states from the dynamical spontaneous electroweak symmetry breaking. 
The framework is implemented in a collider simulation tool especially suited for the study of Goldstone Boson Scattering (GBS), which includes
vector boson scattering (VBS), di-Higgs production via vector boson fusion (VBF) 
and the pair production of other pNGBs via VBF.


\end{abstract}

\maketitle
\tableofcontents


\section{Introduction}

Composite Higgs (CH) models are promising alternatives to the Standard Model (SM), describing the electroweak (EW) symmetry breaking (SB) dynamically via a fermionic condensate, solving the hierarchy problem~\cite{Weinberg:1975gm,Susskind:1978ms,Dimopoulos:1979es}, as well as the little hierarchy between the compositeness scale and the Higgs mass via the vacuum  misalignment mechanism~\cite{Dugan:1984hq} with a SM-like light Higgs boson appearing as a pseudo-Nambu-Goldstone boson (pNGB)~\cite{Kaplan:1983fs,Georgi:1984af}. It also opens several paths towards explaining dark matter with composite states,
unification and SM fermion masses hierarchy via partial compositeness (PC)~\cite{Kaplan:1991dc} with large anomalous dimension of fermionic operators in a near-conformal phase~\cite{Holdom:1981rm}.

The description of CH interactions below the condensation scale is provided by the chiral expansion of the Coleman-Callan-Wess-Zumino (CCWZ) formalism~\cite{Coleman:1969sm,Callan:1969sn}, 
usual in the CH literature (see~\cite{Contino:2010rs,Bellazzini:2014yua,Panico:2015jxa} for reviews).
In CH models an approximate global symmetry $G$ is spontaneously broken to the stability group $H$ giving origin to a set of (p)NGBs.
Besides the 3 exact NGB \emph{eaten} by the weak bosons, CH models contain the Higgs boson as well as other pNGBs in non-minimal scenarios.
Models with an explicit matter content of hyperfermions with PC mechanism via a four-dimensional gauge theory have typically non-minimal cosets~\cite{Barnard:2013zea,Ferretti:2013kya,Vecchi:2015fma,Ferretti:2016upr,Csaki:2017jby,Guan:2019qux}.

The interactions among the (p)NGBs, including the Higgs and longitudinal modes of weak bosons, are fixed by the non-linear symmetry of the CCWZ construction.
These interactions can be observed via precise measurement of SM processes or the production of other pNGBs. 
In this paper we present the CCWZ Lagrangian for a generic CH coset, with a minimal set of operators that allow the description of the gauge interactions and the misalignment of the vacuum.
We include operators at orders $\mO(p^2)$, $\mO(p^4)$ and $\mO(p^2 g^2)$ in the chiral expansion, as well as the topological Wess-Zumino-Witten terms (WZW)~\cite{Wess:1971yu,Witten:1979vv}. 
The presented $\mO(p^4)$ and $\mO(p^2 g^2)$  terms are intended as a minimum self-consistent set of operators that can be used to renormalize one-loop amplitudes in the chiral perturbation framework.

A framework for the automatic simulation of CH processes in collider experiments based on the \feynrules package~\cite{Christensen:2008py,Alloul:2013bka}, Universal Feynrules Output (UFO)~\cite{Degrande:2011ua} and \mg~\cite{Alwall:2011uj} is presented. 
Predictions for (pseudo-)Goldstone boson scattering (GBS) are used as examples. GBS processes give the ultimate test of the dynamical origin of EWSB.
It includes Vector Boson Scattering (VBS), di-Higgs production via Vector Boson Fusion (VBF) and the pair production of extra pNGBs via VBF. 
For the SM processes VBS and di-Higgs via VBF distributions are computed up to $\mO(p^4)$ at tree level.
Predictions for pNGB pair production via VBF for $SU(4)/Sp(4)$ and $SU(5)/SO(5)$ are also produced at $\mO(p^2)$ at tree level.

The paper is organized as follows. 
In \sec{sec:lo} we provide the general CCWZ framework and the contruction of the lowest order Lagrangian, including the WZW terms. 
In \sec{sec:nlo} we extend the formalism to include operators of $\mO(p^4)$ and $\mO(p^2g^2)$.
In \sec{sec:gbs} we show some predictions as example.
We conclude in \sec{sec:conclusion} with prospects for the future. 

\section{CCWZ formalism for CH models to lowest order}
\label{sec:lo}

In this section we repeat the CCWZ formalism for CH models in order to fix the notation.
The basic idea of the CCWZ construction is to define objects which are invariant under the global symmetry $G$, that are spontaneously broken to the subgroup $H$ via a ``gauge" choice.

The generators of $G$, $T^A$, $A=1,\cdots,d_G$, are normalized with the convention $\langle T^A T^B \rangle = \frac{1}{2}\delta^{AB}$. $\langle \rangle$ is the trace.
They are split in the unbroken generators belonging to $H$, $T^a\equiv \tilde{T}^a\equiv S^a$, $a=1,\cdots,d_H$ and the broken ones that belong to the coset $G/H$, $T^{d_H+\ha} \equiv \hat{T}^{\ha} \equiv X^{\ha}$, $\ha=1,\cdots,d_G-d_H$.
For the cosets which can be realized with fermion condensates $SU(2N)/SO(2N)$, $SU(2N)/Sp(2N)$ and $SU(N)_L\times SU(N)_R/SU(N)_V$~\cite{Peskin:1980gc,Preskill:1980mz}, we can get the generators from the equations
\begin{equation}
S^a\ez+\ez S^{aT}=0, \quad X^{\ha}\ez - \ez X^{\ha T}=0,
\end{equation}
with the $\ez$ a (anti-)symmetric matrix in the (pseudo-)real rep. $SU(N)/SO(N)$ ($SU(N)/Sp(N)$) case. 
We choose explicitly
\begin{equation}
\ez= 
\begin{cases}
\id_{2N},\quad \text{for } SU(2N)/SO(2N)  \\
 \left( 
 \begin{array}{ccc}
			& -\epsilon_N &  \\
\epsilon_N 	&  			&  \\
			&  			& 1  
\end{array} \right), \quad \text{for } SU(2N+1)/SO(2N+1)  \\
 \left( 
\begin{array}{cc}
		& -\id_N  \\
\id_N  	&  
\end{array} \right), \quad \text{for } SU(2N)/Sp(2N) \\
 \left( 
\begin{array}{cc}
& \id_N  \\
\id_N  	&  
\end{array} \right), \quad \text{for } SU(N)_L\times SU(N)_R/SU(N)_V
\end{cases}
\end{equation}
$1_N$ is the $N$-dimension identity matrix and $\epsilon_N$ is the fully anti-symmetric $N$-dimension matrix with entries given by the Levi-Civita symbol with $\epsilon_{12\dots N}=1$.

In the $SU(N)^2/SU(N)$ case, we use
\begin{equation}
T^A=
 \left( 
\begin{array}{cc}
T_L^A 	&   \\
		&   -T_R^{A T}
\end{array} \right)
,\quad T_{L(R)}^A=\begin{cases} \lambda^A (\id_N), \, A=1,\cdots, \text{dim}(SU(N)) \\
\id_N (\lambda^A), \, A = \text{dim}(SU(N))+1,\cdots,2 \text{dim}(SU(N))\end{cases}
\end{equation}
With this choice for $T^A$ and $\ez$ we get the unbroken $T_L=T_R$ and the broken $T_L=-T_R$ generators as block diagonal matrices:
\begin{equation}
S^a =\text{diag}(\lambda^a,-\lambda^{Ta}),\quad 
X^a = \text{diag}(\lambda^a,\lambda^{Ta}) 
\end{equation} 
where $\lambda^a$ are the Gell-Mann matrices generators of $SU(N)$. The advantage of this notation is that all subsequent discussion can be made general.
For the minimal CH  coset $SO(5)/SO(4)$ we use~\cite{Contino:2011np}
\begin{align}
X^{\ha} &= -\frac{i}{\sqrt{2}}(\delta^{\ha i}\delta^{5 j} - \delta^{\ha j}\delta^{5 i}), \\
S^a_{ij} &=  -\frac{i}{2}(\frac{i}{2}\epsilon^{abc}(\delta^{b i}\delta^{c j} - \delta^{b j}\delta^{c i}) + (\delta^{a i}\delta^{4 j} - \delta^{a j}\delta^{4 i})  ),\\
S^{a+3}_{ij} &=  -\frac{i}{2}(\frac{i}{2}\epsilon^{abc}(\delta^{b i}\delta^{c j} - \delta^{b j}\delta^{c i}) - (\delta^{a i}\delta^{4 j} - \delta^{a j}\delta^{4 i})  ),
\end{align}
with $a=1,2,3$.

An element of $G$ and of $H$ are $g=e^{i\alpha^A T^A}$ and $h=e^{i\sigma^a S^a}$ respectively.
Every element of $G$ can be parameterized by $g=\xi(\pi) h$
with $\xi$ an element of the coset $G/H$ 
\begin{equation}
\xi(\pi)=e^{i \pi^A X^A}
\end{equation} 
The action of $g\in G$ in $\xi(\pi)$ is still an element $g'\in G$ and can be parametrized in the same form
$g'\equiv g\xi(\pi)=\xi(\pi')h(\pi,g)$
defining the non-linear character of $\xi(\pi)$ transformation
\begin{equation}
\xi(\pi)\to \xi(\pi')=g \xi(\pi) h(\pi,g)^{-1}\,.
\end{equation}
It is also useful to define 
\begin{equation}
\Sigma \equiv \xi\ez \xi^T\to g\Sigma g^T \,.
\end{equation}
which follows from 
$\ez h = \ez e^{\ii \alpha^a S^a}= e^{-\ii \alpha^a S^{a T}}\ez=h^*\ez$.

We can now embed the left $ T_L^i$ and right $ T_R^i$ generators of the custodial subgroup $SU(2)_L\times SU(2)_R\subset H$ as unbroken generators. This can be defined as any 3+3 generators that obey the SU(2) Lie algebra
\begin{equation}
[T_{L,R}^a,T_{L,R}^b]=\ii \epsilon^{abc} T_{L,R}^c
,\quad
[T_L^a,T_R^b]=0\,.
\end{equation}
We will eventually \emph{gauge} $SU(2)_L$ and the hypercharge as the third component of $SU(2)_R$. We will not consider cases where $H$ has no custodial group as subgroup. 
We keep the structure constant $\epsilon^{abc}$ as in the SM, in order to conserve the usual values of the gauge couplings $g,g'$. This enforces us to have in general a different normalization for the generators 
\begin{equation}
\Tr[T_{L,R}^aT_{L,R}^b]= t\delta^{ab}\,.
\end{equation}
In this manuscript we consider only condensates in the EW sector and will not discuss QCD charged condensates, even though they are important in CH models and in top quark partial compositeness contruction in particular~\cite{Ferretti:2013kya,Ferretti:2016upr}.

We can now identify the Higgs doublet $\Phi$ as a NGB transforming as a bi-doublet of  $\SU(2)_L\times\SU(2)_R$ just defined. The neutral component $\tilde{h}$ (with associated generator  $X_h$) acquires a vacuum expectation value (vev) $\langle \tilde{h}\rangle$ to break EW symmetry. Similarly, other NGB can acquire a vev.   
The $\xi$ object can thus be written as
\begin{equation}
\xi = \Omega e^{i\Pi/f}  \,,\quad \text{with   } \Pi= N \pi^a  X^a \,.
\end{equation}
The fields $\pi^a$ are free of tadpoles, since the form of $\Omega$ is provided by the minimization of the potential. $N$ is a normalization factor.
We note that the new fields are not simply related by $\tilde{h}=\langle \tilde{h}\rangle +h$.
$\Omega$ is given by the exponentiation of the vev
\begin{equation}
\Omega = e^{i\langle\tilde{\Pi}\rangle/f} \Rightarrow e^{\ii N X_h \langle \tilde{h}\rangle/f} = e^{\ii N X_h \theta'}
\end{equation}
where the arrow stands for the case where  only the Higgs gets a vev and $X_h$ is the respective generator.

To identify the Higgs and construct the $\Omega$ matrix we notice that it is the real part of a bi-doublet of $\SU(2)_L\times \SU(2)_R$.
Under $h\in H$ transformation, we have 
\begin{equation}
\xi\to g\xi \hat{h}^{-1}(g,\pi) \Rightarrow \xi \xrightarrow{g=h} h\xi h^{-1}
\end{equation}
and so $\Pi\to h\Pi h^{-1}$ and by expanding we can get the eigensystem of specific generators $[T,\Pi]=\lambda(T)\Pi$.
The neutral part of the bidoublet $H^0 =h \pm \ii \pi^3$ has eigenvalues $\lambda(T_L^3)= - \lambda(T_R^3) = \pm 1/2$ and from that we can identify the Higgs with the real component
\footnote{Alternatively, we can use the fact that the Higgs is a singlet of $\SU(2)_V$ while $\pi^3$ is part of a triplet, and get rid of the triplet component.}.
After identifying the Higgs we can exponentiate it to get the $\Omega$ matrix.
We provide the explicit form of $X_h$ and $\Omega$ for the ``minimal cosets" $SU(4)/Sp(4)$, $SU(5)/SO(5)$ and $SU(4)\times SU(4)/SU(4)$ in \app{sec:group}.

In general other pNGB might acquire a vev, and in this case the whole vacuum has to be exponetiated to get the $\Omega$ matrix. 
We will not explore this possibility in this paper, although it can have interesting phenomenological implications.

The construction of the Lagrangian in the condensate phase follows the usual chiral Lagrangian prescription, with the definition of some basic objects,
\begin{equation}
	\omega_{\mu}  =  \xi^\dag \nabla_{\mu} \xi\,,\quad
\nabla_\mu \xi = ( \partial_\mu -i j_\mu ) \xi ,\quad j_\mu = v_\mu^a S^a + a_\mu^{\ha} X^{\ha}
\end{equation}
which  transform under $G$ as 
\begin{equation}
\omega_\mu \to h\omega_\mu h^\dagger + h\partial_\mu h^\dagger\,,\quad 
j_\mu\to g j_\mu g^\dagger+ i g\partial_\mu g^\dagger.
\end{equation}
$\omega_\mu$ can be further decomposed into projections to the unbroken and broken directions as
\begin{align}
x_\mu &= 2\langle X^{\ha} \omega_\mu \rangle X^{\ha}, \quad  x_\mu \to  h x_\mu h^\dagger, \label{eq:xmu}\\
s_\mu &= 2\langle S^a \omega_\mu \rangle S^a, \quad  s_\mu \to  h s_\mu h^\dagger +  h \partial_\mu h^\dagger\,. \label{eq:smu}
\end{align}
We can also define a field-strength that transforms homogeneously
\begin{equation}
s_{\mu\nu} = \partial_\mu s_\nu - \partial_\nu s_\mu + [s_\mu,s_\nu] \to h s_{\mu\nu} h^\dagger \,.
\end{equation}

The incorporation of propagating EW bosons is similar to the incorporation of virtual photons in the QCD chiral theory~\cite{Urech:1994hd}.
The vector current is simply the gauge interactions. We can thus turn off the fictitious gauge bosons and keep only the real EW bosons
\begin{eqnarray}
 \label{eq:vectors0}
j_\mu &=& g {\bf \widetilde{W}}_\mu + g^\prime {\bf B_\mu} \\
{\bf B_\mu}&=&B_\mu\ T_R^3, \quad  {\bf\widetilde{ W}_\mu}=\sum_{a=1}^{3} \widetilde{W}_\mu^a\  T_L^a, 
\end{eqnarray}
where $\widetilde{W}^k_\mu$ ($k=1,\,2,\,3$) and $B_\mu$ are the elementary electroweak gauge bosons associated with the $SU(2)_L$ and $U(1)$ hypercharge groups. 

Hyperfermion current masses can be parameterized via the following spurion transforming homogeneously
\begin{equation}
 \chi = f^2\xi^\dagger \mM\xi^*\ez \to h\chi h^\dagger
\end{equation}
where $\mM$ is the hyperfermion mass matrix with dimensionless coefficients of order $m/f$ where $m$ are hyperquark current masses.
We can also define a hermitian and an anti-hermitian combination
\begin{equation}
\chi_\pm = \frac{1}{2}(\chi\pm \chi^\dagger)\,.
\end{equation}

 The fact that the condensate are EW charged also require the definition of gauge spurions
\begin{equation}
 \Gamma^{g,i} = \xi^\dagger T_L^i\xi\to h\Gamma^{g,i}  h^\dagger,\quad \Gamma^{g'} = \xi^\dagger T_R^3\xi\to h\Gamma^{g'}  h^\dagger.
 \label{eq:gauge_spurion}
\end{equation}


With the basic objects defined above and a power counting scheme~\cite{Manohar:1983md,Gavela:2016bzc}, we can construct the lowest order Lagrangian.
\begin{align}
\mL_{LO} &= \frac{f^2}{N^2}\langle x^\mu x_\mu + \chi_+\rangle -\frac{1}{4}W_{\mu\nu}^aW^{a,\mu\nu} -\frac{1}{4}B_{\mu\nu}B^{\mu\nu} \nonumber \\
 &+ C_g f^4 g^2 \langle  \Gamma^{g,i}\ez\Gamma^{g,iT}\ez \rangle +C_g' f^4 g'^2 \langle  \Gamma^{g'}\ez\Gamma^{g'T}\ez \rangle .
\label{eq:L2}
\end{align}

The coefficients $C_{g,g'}$ are order 1. The corresponding $C$ in QCD with the addition of the photon, that gives the mass difference between the charged and neutral pions is $~0.8$~\cite{Urech:1994hd}.


The Lagrangian \label{eq:L2} sets the LO relation between the NGB decay constant and the EW scale $v=246\GeV$,
\begin{equation}
v=f\sin\theta\,.
\end{equation}

The SM fermion kinetic term is provided by the usual SM covariant term, while their Yukawa couplings are model dependent.
The usual approach in dynamical EW symmetry breaking is the existence of another type of interaction at the flavor scale $\Lambda_F\sim 10^4\TeV$, which induces 4-fermion interactions containing both the hyper fermions and SM fermions, which at low energy after condensation give origin to the SM fermion masses and their Yukawa couplings with the pNGBs.
The form of these Yukawa terms depends on the physics at the flavor scale but can be described by spurionic fields transforming under $G$. 
For sake of simplicity we follow \cite{Alanne:2018wtp} and concentrate here on spurions transforming in the fundamental (F), two-index symmetric (S) or anti-symmetric (A) and the adjoint (Adj) representations.
Each one transforms as
\begin{eqnarray}
\Xi_F\to g\Xi_F,\quad \Xi_{S/A}\to g\Xi_{S/A}g^T,\quad \Xi_{Adj}\to g\Xi_{Adj}g^\dagger.
\end{eqnarray}
The SM fermions can be embedded in these representations after identifying the object that transforms according to their SM quantum numbers. 
We can then define the fermion fields in 2-index irreps and fundamental transforming as 
\begin{align}
\psi &= \begin{cases} \xi^\dagger \Xi_{A/S}\xi^*\ez  \\
\xi^\dagger \Xi_{Adj}\xi \end{cases} \to h\psi h^\dagger, \\
\psi^F&=\xi^\dagger \Xi_F \to h \psi
\label{eq:fermions}
\end{align}
The Yukawa terms follow from the bilinear terms like $\bar\psi \psi$.
The Lagrangian describing SM fermions at lowest order is thus
\begin{align}
\mL_{\psi} &=  \bar{Q} i\slashed D Q + \bar{L} i\slashed D L  + \bar{u} i\slashed D u + \bar{d} i\slashed D d + \bar{e} i\slashed D e\\ 
&- \frac{f}{4\pi} \left( Y^1_u \langle \bar{\psi}_Q \psi_u \rangle + Y^1_d \langle\bar{\psi}_Q \psi_d \rangle + Y^1_e \langle \bar{\psi}_L \psi_e \rangle + Y^1_\nu \langle \bar{\psi}_L \psi_\nu \rangle + \hc \right)\\
&- \frac{f}{4\pi}\left(  Y^2_u\langle \bar{\psi}_Q\rangle \langle \psi_u \rangle + Y^2_d \langle \bar{\psi}_Q\rangle \langle  \psi_d \rangle + Y^2_e \langle \bar{\psi}_L\rangle \langle \psi_e \rangle 
   + Y^2_\nu \langle \bar{\psi}_L \rangle  \langle \psi_\nu \rangle + \hc \right)\\
&- \frac{f}{4\pi} \left(  Y^F_u \bar{\psi}^F_Q \psi^F_u  + Y^F_d  \bar{\psi}^F_Q \psi^F_d + Y^F_e \langle \bar{\psi}^F_L \psi^F_e
	+ Y^F_\nu \bar{\psi}^F_L  \psi^F_\nu  + \hc \right)
\end{align}
where $Q$ are left-handed quarks, $L$ are the left handed, $u,d$ the up and down right handed quarks and $e$ the right-handed electron. 
All should be understood as sum over flavour space.  $Y_{u,d,e,\nu}$ are matrices in flavour space with coefficients related to the pre-Yukawas from the flavor scale, and $\psi_f$ are the spurion embedding the corresponding fermion $f$ and defined as \ref{eq:fermions}.
Explicit embeddings can be found in~\cite{Golterman:2017vdj, Alanne:2018wtp}.

The set of terms in the effective potential that needs to be added to renormalize SM fermion loops can be derived from the interactions above. 
We leave this for future work and will be concerned with only pNGBs and EW bosons in loops.

The topological Wess-Zumino-Witten terms~\cite{Wess:1971yu,Witten:1983tw} are universal and model independent, besides being phenomenologically relevant in particular for the description of bosonic decays of pNGBs. 
They are given in differential form by~\cite{Kaymakcalan:1983qq,Ferretti:2016upr,Brauner:2018zwr}
\begin{align}
\nonumber
S_{\mathrm WZW} \supset \frac{i \rm{dim}(\psi)}{48\pi^2}\int & \langle\bigg(dA_L A_L dU U^\dagger + A_L dA_L dU U^\dagger + dA_R A_R U^\dagger dU + A_R dA_R U^\dagger dU\\
& - dA_L dU A_R U^\dagger + dA_R dU^\dagger A_L U \bigg)\rangle.
\end{align}
For $\rm SU(4)/Sp(4)$ and  $\rm SU(5)/SO(5)$: $A_L = A$, $A_R = -A^T = -\epsilon A \epsilon$, while for $\rm SU(4)\times SU(4)^\prime/SU(4)_D$: $A_L = A_R = A$.
Expanding in the first order in the pNGB and integrating by parts yields
\begin{equation}
\mL_{WZW}= \frac{\rm{dim}(\psi)}{48\pi^2 f}\langle 2F_{\mu\nu}\widetilde{F}^{\mu\nu} (\Omega \Pi\Omega^\dagger + \Omega^\dagger \Pi\Omega) 
	+ \Omega^\dagger F_{\mu\nu} \Omega\Pi \Omega \widetilde{F}^{\mu\nu}\Omega^\dagger 
	+ \Omega F_{\mu\nu} \Omega^\dagger \Pi \Omega^\dagger \widetilde{F}^{\mu\nu}\Omega \rangle \,.
\end{equation}
$\widetilde{F}^{\mu\nu}\equiv \frac{1}{2}\epsilon^{\mu\nu\rho\sigma}{F}_{\rho\sigma}$.

\section{The next-to-leading order Lagrangian}
\label{sec:nlo}

To construct the $\ord(p^4)$ Lagrangian we need to define some other objects.
The vector field strength
\begin{equation}
j_{\mu\nu}=\partial_\mu j_\nu - \partial_\nu j_\mu + [j_\mu, j_\nu] \to g j_{\mu\nu} g^\dagger
,\quad f_{\mu\nu} = \xi^\dagger j_{\mu\nu}\xi \to hf_{\mu\nu}h^\dagger
\end{equation}
and its projections $f_{\mu\nu}=\tilde{f}_{\mu\nu}+\hat{f}_{\mu\nu}$
as in \eqs{eq:xmu}{eq:smu}.

The Leutwyler-Gasser terms~\cite{Gasser:1983yg,Gasser:1984gg} at NLO, worked out for a generic coset in ~\cite{Bijnens:2009qm}, are given by
\begin{align}
\mL_{p^4} &= \frac{1}{16\pi^2}\big\{ L_0 \langle x^\mu x^\nu x_\mu x_\nu \rangle +  L_1 \langle x^\mu x_\mu \rangle  \langle x^\nu x_\nu \rangle + 
			L_2 \langle x^\mu x^\nu \rangle  \langle x_\mu x_\nu \rangle + L_3\langle x^\mu x_\mu x^\nu x_\nu \rangle \nonumber \\
     &+ L_4 \langle x^\mu x_\mu \rangle \langle \chi_+ \rangle + L_5 \langle x^\mu x_\mu \chi_+ \rangle  
     + L_6 \langle \chi_+ \rangle ^2 + L_7 \langle \chi_-\rangle ^2 + \frac{1}{2}L_8\langle \chi_+^2 +\chi_-^2 \rangle \nonumber \\
     &-\ii  L_9 \langle \tilde{f}_{\mu\nu}x^\mu x^\nu \rangle + \frac{1}{4} L_{10} \langle \tilde{f}_{\mu\nu}^2 - \hat{f}_{\mu\nu}^2 \rangle \big\}
\label{eq:L4}
\end{align}
We add a $1/(4\pi)^2$ from naive dimension analysis (NDA) power counting~\cite{Manohar:1983md} and so the coefficients are expected to be $\mO(1)$.

Once gauge interactions are turned on, we need to add the gauge spurions $\Gamma$ (\eq{eq:gauge_spurion}) and to define their covariant derivative
\begin{equation}
D_\mu\Gamma=\partial_\mu\Gamma + [s_\mu, \Gamma]\to h D_\mu\Gamma h^\dagger \,.
\end{equation}
Following ~\cite{Urech:1994hd} we obtain the corresponding terms for the Lagrangian,
\begin{align}
\mL_{g^2 p^2} &= \frac{g^2 f^2}{16\pi^2} \big\{ K_1 \langle x_\mu x^\mu \rangle\langle \Gamma^2\rangle
			+ K_2 \langle x_\mu x^\mu \rangle\langle \Gamma^{g,i}\ez\Gamma^{g,iT}\ez \rangle
			+ K_3 \langle x_\mu \Gamma \rangle\langle x^\mu \Gamma\rangle \nonumber \\
			&+ K_5 \langle x_\mu x^\mu  \Gamma^2\rangle
			+ K_6 (\langle x_\mu x^\mu \Gamma^{g,i}\ez\Gamma^{g,iT}\ez \rangle + \hc )
			+ K_7 \langle \chi_+ \rangle \langle \Gamma^2\rangle \nonumber \\			
			&+ K_8 \langle \chi_+ \rangle \langle \Gamma^{g,i}\ez\Gamma^{g,iT}\ez\rangle
			+ K_9  \langle \chi_+  \Gamma^2\rangle
			+ K_{10} (\langle \chi_+  \Gamma^{g,i}\ez\Gamma^{g,iT}\ez\rangle +\hc) \nonumber \\			
			&+ K_{11} (\langle \chi_-  \Gamma^{g,i}\ez\Gamma^{g,iT}\ez\rangle +\hc)
			+ K_{12}  (\langle x^\mu D_\mu\Gamma^{g,i}\ez\Gamma^{g,iT}\ez \rangle +\hc) \nonumber \\	
			& + K_{13} D_\mu\Gamma^{g,i}\ez (D^\mu\Gamma^{g,i})^T\ez 			
			+ K_{14} D_\mu\Gamma^{g,i} D^\mu\Gamma^{g,iT} \big\}
\label{eq:Lg2p2}
\end{align}
Similar terms replacing $g$ with $g'$ should also be added.

\section{Towards predictions: GBS at the LHC}
\label{sec:gbs}

The framework described in sections \ref{sec:lo} and \ref{sec:nlo} is implemented for some cases of interest in the the \textsc{FeynRules} package~\cite{Alloul:2013bka, Christensen:2008py}.
The available models are\footnote{The \textsc{FeynRules} and UFO models can be retrieved from the High Energy Model Database \url{https://hepmdb.soton.ac.uk/hepmdb:0223.0338}.}: 
\begin{itemize}
	\item the minimal CH (MCH) model based on the coset $SO(5)/SO(4)$, including terms of $\mO(p^4)$, $\mL_{p^4}$ (topological terms in $L_{WZW}$ vanish in this coset).
	\item the $SU(4)/Sp(4)$ model including topological terms $L_{WZW}$, suited for the phenomenological study of the extra singlet pNGB and its bosonic decays.
	\item the $SU(5)/SO(5)$ model including topological terms $L_{WZW}$, suited for the study of the 14 (p)NGBs and their bosonic decays.
\end{itemize}

To illustrate the power of the developed tools we discuss the effect of $\mO(p^4)$ terms at tree-level in di-Higgs production and VBS and the pair production of extra pNGBs in the two implemented non-minimal models.
All simulation is performed with the export of the \feynrules model to the Universal FeynRules Output (UFO)~\cite{Degrande:2011ua}, that can be imported in the \mg program~\cite{Alwall:2011uj} for simulations of proton-proton scattering.
The simulations are performed  at 14 TeV in the proton-proton center-of-mass energy with the NNPDF 3.1 NLO LUXQED parton distribution function (PDF) set with $\alpha_s(\mu)=0.118$~\cite{Ball:2013hta}.

\subsection{Di-Higgs via VBF and VBS at $\mO(p^4)$ and tree-level}

To study the Higgs and weak boson physics at high energy we neglect possible extra pNGBs, having in mind that the Higgs and EW bosons obey universal relations.
We also neglect the the spurion contributions, $\chi\to 0$ in $\mL_4$ and the $\mL_{g^2p^2}$ (\eq{eq:Lg2p2}) as a first approximation. 
As pointed out in ~\cite{Liu:2018qtb}, the modification in these interactions can be well approximated by the MCH SO(5)/SO(4) description.
In SO(5)/SO(4) the $L_0$ and $L_3$ terms are redundant and can be rewritten in terms of $L_1$, $L_2$. 
An interesting particularity in the $SO(5)/SO(4)$ coset is that the stability group $H$ is not simple, $SO(4)\sim SU(2)_L\times SU(2)_R$. 
This fact allow us to further break the projections into the unbroken generators into two independent terms. For example $\tilde{f}_{\mu\nu}$ can be broken in
\begin{align}
O_L & \equiv  O_L^a S^a = 2\langle S^a \tilde{O} \rangle S^a, \, a=1,2,3 \quad \\
O_R & \equiv  O_R^a S^a = 2\langle S^a \tilde{O} \rangle S^a, \, a=4,5,6 \,.
\end{align}
Therefore, we use a different basis for the operators
\begin{align}
\mL^{MCH}_4 &= \frac{1}{16\pi^2}\big\{ L_1 \langle x^\mu x_\mu \rangle  \langle x^\nu x_\nu \rangle + 
			L_2 \langle x^\mu x^\nu \rangle  \langle x_\mu x_\nu \rangle 
			+ L_3^-\langle s^{\mu\nu}_L s_{\mu\nu, L} - s^{\mu\nu}_R s_{\mu\nu, R} \rangle\nonumber \\    
     &-\ii  L_9^+ \langle (f^{\mu\nu}_L + f^{\mu\nu}_R) x_\mu x_\nu \rangle 
      -\ii  L_9^- \langle (f^{\mu\nu}_L - f^{\mu\nu}_R) x_\mu x_\nu \rangle \nonumber \\     
     &+ \frac{1}{4} L_{10}^+ \langle \hat{f}_{\mu\nu}\hat{f}^{\mu\nu} \rangle 
      + \frac{1}{4} L_{10}^- \langle f_L^{\mu\nu}f_{\mu\nu,L} - f_R^{\mu\nu}f_{\mu\nu,R} \rangle \big\}
  \label{eq:LagMCH}
\end{align}

The coefficients carrying a minus sign break the $Z_2$ symmetry $R\leftrightarrow L$.
Notice that in several models the $Z_2$ symmetry is present in the strong theory and is only softly broken. For instance $O(4)$ is a subgroup of $Sp(4)$.
However, the breaking of $Z_2$ can be achieved in other cosets by gauge spurions $\Gamma^{g,g'}$ and the appropriate choice of the hyperfermion spurion $\chi$.
The Lagrangian is equivalent to \cite{Contino:2011np,Liu:2018qtb}, using the NDA power counting $m_\rho=4\pi f$ and the identification,
\begin{equation}
c_1=L_1,\,c_2=L_2,\,c_3=L_3^-,\,c_4^\pm=L_9^\pm,\,c_5^\pm=L_{10}^\pm\,.
\end{equation}

The implementation in \feynrules requires the fields to have canonically normalized kinetic terms and diagonal masses. 
The higher dimension operators induce corrections to the kinetic terms, as
\begin{equation}
\mL_4 \supset -\frac{1}{4}B_{\mu\nu}B^{\mu\nu}[1+X_B] -\frac{1}{4}W^i_{\mu\nu}W^{i\mu\nu}[1+X_W] -\frac{1}{4}W^3_{\mu\nu}B^{\mu\nu}X_{BW}
\end{equation}
 and one must thus redefine the fields to fulfill these requirements.
This is done by the following replacements:
\begin{align}
W^i_\mu &\to  W^i_\mu(1+X_W) \\
A_\mu &\to  A_\mu (1 + X_W s_{W,0}^2 + X_B c_{W,0}^2 + 2 s_{W,0} c_{W,0} X_{BW})+ \nonumber \\
&    Z_\mu (s_{W,0} c_{W,0} (X_W-X_B) + (c_{W,0}^2-s_{W,0}^2) X_{BW} + \delta s_W/c_{W,0}), \\
Z_\mu &\to  Z_\mu (1 + X_W c_{W,0}^2 + X_B s_{W,0}^2 - 2 s_{W,0} c_{W,0} X_{BW})+ \nonumber \\
&    A_\mu (s_{W,0} c_{W,0} (X_W-X_B) + (c_{W,0}^2-s_{W,0}^2)X_{BW} - \delta s_W/cw).
\end{align}
Also the gauge couplings have to be modified,
\begin{align}
g_2 &\to g_2/(1+X_W) \sim g_2(1-X_W) \\
g_1 &\to g_1/(1+X_B) \sim g_1(1-X_B) \\
s_{W,0}^2 &= (1-M_W^2/M_Z^2) \\
s_W &= s_{W,0}(1 + \delta s_W^2) \\
\delta s_W &= X_{BW} c_{W,0} (1-2s_{W,0}^2)
\end{align}
The exact form of the mixing coefficients in terms of the Lagrangian coefficients are
\begin{align}
X_W &= \frac{e^2}{8s_{W,0}^2}[6L_9^+ + L_{10}^+ + 4(4L_3^+ + 2L_9^- + L_{10}^-)\ct + (2L_{9}^+ - L_{10}^+)\ctt], \\
X_B &= \frac{e^2}{8c_{W,0}^2}[6L_{9}^+ + L_{10}^+ - 4(4L_{3}^- + 2L_{9}^- + L_{10}^-)\ct + (2L_{9}^+ - L_{10}^+)\ctt], \\
X_{WB} &= \frac{e^2}{4 c_{W,0} s_{W,0}}(2L_{9}^+ - L_{10}^+)\st^2
\end{align}

\begin{figure}
\includegraphics[width=0.45\textwidth]{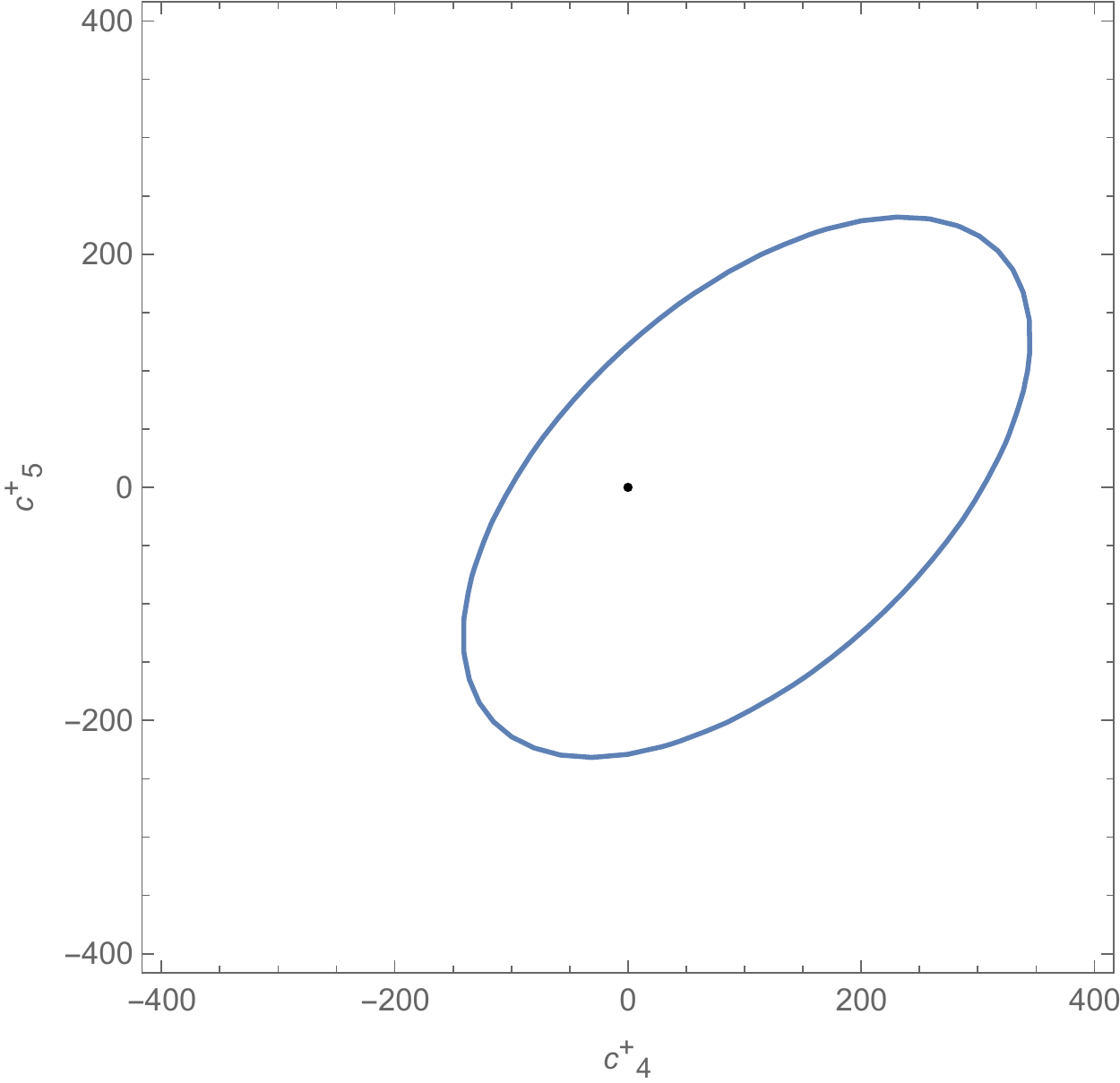}
\caption{Bounds on $L_9^+$ and $L_{10}^+$ assuming $Z_2$ symmetry $L_9^-=L_3=0$ for $\theta=0.3$ ($f=832\GeV$). The bounds scale with $f_{NEW}^2/f^2$ becoming less constraining for large $f_{NEW}$.}
\label{fig:aTGC}
\end{figure}

The coefficients $L_3^-$, $L_9^\pm$ and $L_{10}^+$  contribute to anomalous trilinear gauge couplings (aTGC)
\begin{align*}
\delta g_{1,z}=\left(\frac{m_Z^2}{(4\pi f)^2}\right)(-2L_3^-+L_9^-)\cos\theta +L_9^+,
\quad \delta \kappa_\gamma = \left(\frac{m_W^2}{(4\pi f)^2}\right)(-4(L_9^+ -2 L_{10}^+))
\end{align*}
and $\delta\kappa_z=\delta g_{1,z}-\frac{s_W^2}{c_W^2}\delta\kappa_\gamma$.
Using the fit done in ~\cite{1508.00581} we can put bounds on those coefficients. But only with these modifications it is not possible to lift the degeneracy since there can be cancellations. 
On the other hand, if we assume $Z_2$ symmetry ($L_9^-=L_3=0$), a bound can be set on $L_9^+$ and $L_{10}^+$, as shown in \fig{fig:aTGC}. We use $\theta=0.3$ ($f=832\GeV$).
The bounds are typically weak, especially considering order 1 coefficients 
(in QCD for instance $L_i\lesssim 1$~\cite{Ecker:1994gg}).

The Lagrangian \ref{eq:LagMCH} contains also anomalous quartic gauge couplings (aQGC). 
We use the conventions of ~\cite{1604.03555} to define the effective couplings, 
\begin{align}
\mO_{WW,1}&=W^{+\mu}W^-_\mu W^{+\nu} W^-_\nu, & \mO_{WW,2}&=W^{+\mu}W^{-\nu} W^{+}_{\mu} W^-_\nu, \\
\mO_{WZ,1}&=W^{+\mu}W^-_\mu Z^{\nu} Z_\nu, & \mO_{WZ,2}&=W^{+\mu}W^{-\nu} Z_{\mu} Z_\nu, \\
\mO_{ZZ,1}&=Z^{\mu}Z^\mu Z^{\nu} Z_\nu, &  &\\
\mO_{WA,1}&=W^{+\mu}W^-_\mu A^{\nu} A_\nu, & \mO_{WA,2}&=W^{+\mu}W^{-\nu} A_{\mu} A_\nu, \\
\mO_{AZ,1}&=W^{+\mu}W^-_\mu A^{\nu} Z_\nu, & \mO_{AZ,2}&=W^{+\mu}W^{-\nu} A_{\mu} Z_\nu + \hc
\end{align}
Only zero derivative terms are generated at this order.
The operators that contribute to aTGC ($L_3^-$, $L_9^\pm$ and $L_{10}^+$) contribute exclusively to a modification to quartic coupling with the same Lorentz structure of the SM, while the so-called genuine quartic coupling operators $L_1$ and $L_2$ contribute with a different Lorentz structure (and $L_{10}^-$ do not contribute at all). 
The full QGC Lagrangian is
\begin{align}
\mL_{QGC}&=-g^2(\mO_{WW,1}-\mO_{WW_2})\left[1 + \frac{m_W^2}{16\pi^2 f^2}\left( - 8 \cos\theta L_3^- + L_4^+ + \cos\theta L_4^-  \right)\right] \nonumber \\
		&+ \frac{m_W^4}{\pi^2 f^4}\left[ 2L_1\mO_{WW,1} + L_2(\mO_{WW,1}+\mO_{WW,2}) \right] \nonumber\\
		&+g^2c_W^2(\mO_{WZ,1}-\mO_{WZ_2})\left[1 + \frac{m_Z^2}{16\pi^2 f^2}\left( 8 \cos\theta L_3^- - (1-4s_W^4)L_4^+ + \cos\theta L_4^- +s_W^4 L_5^+  \right)\right] \nonumber \\
		&+ \frac{m_W^2m_Z^2}{\pi^2 f^4}\left( 2L_1\mO_{WZ,1} + L_2\mO_{WZ,2} \right) \nonumber\\
		&-e^2\frac{c_W}{s_W}(\mO_{AZ,1}-\mO_{AZ_2})\left[1 + \frac{m_Z^2}{16\pi^2 f^2}\left(-16 \cos\theta L_3^- + 2(4s_W^4-4s_W^2-1)L_4^+ - \cos\theta L_4^- + s_W^4 L_5^+  \right)\right] \nonumber \\
		&+ \frac{m_W^2m_Z^2}{\pi^2 f^4}\left( 2L_1\mO_{WZ,1} + L_2\mO_{WZ,2} \right) \nonumber\\		
		&+\frac{m_Z^4}{\pi^2 f^4}2\left( L_1\mO_{ZZ,1} + L_2\mO_{ZZ,2} \right) \nonumber\\		
		&+e^2(\mO_{WA,1}-\mO_{WA_2})\left[1 + \frac{e^2v^2}{16\pi^2 f^2}\left( 2L_4^+ - L_5^+  \right)\right] 	
\end{align}

With the model implementation and a better understanding of its modifications to the SM interactions we perform the simulation of di-Higgs production via VBF and VBS at the LHC. 

\begin{figure}[htbp]
  	\centering
  	\includegraphics[width=0.25\textwidth]{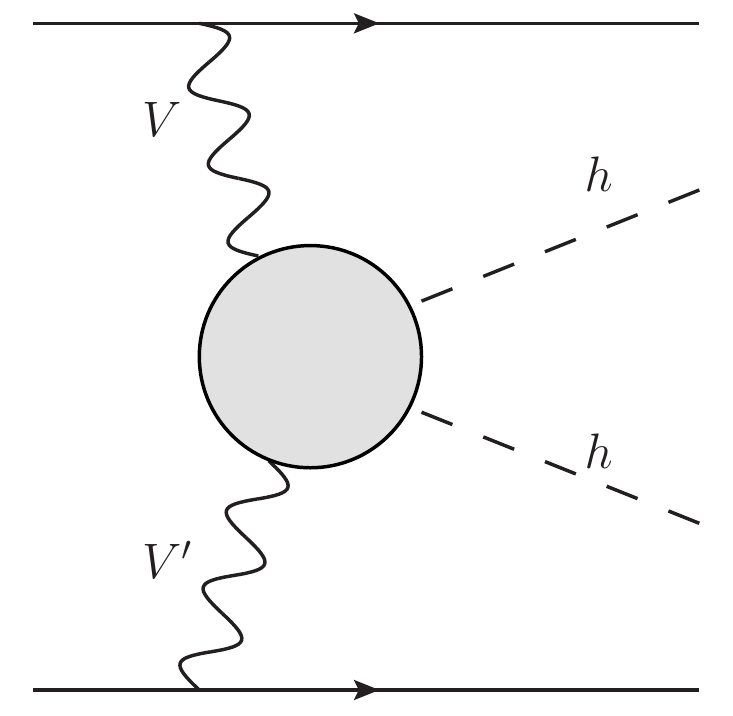}
  	\hspace{2cm}
  	\includegraphics[width=0.25\textwidth]{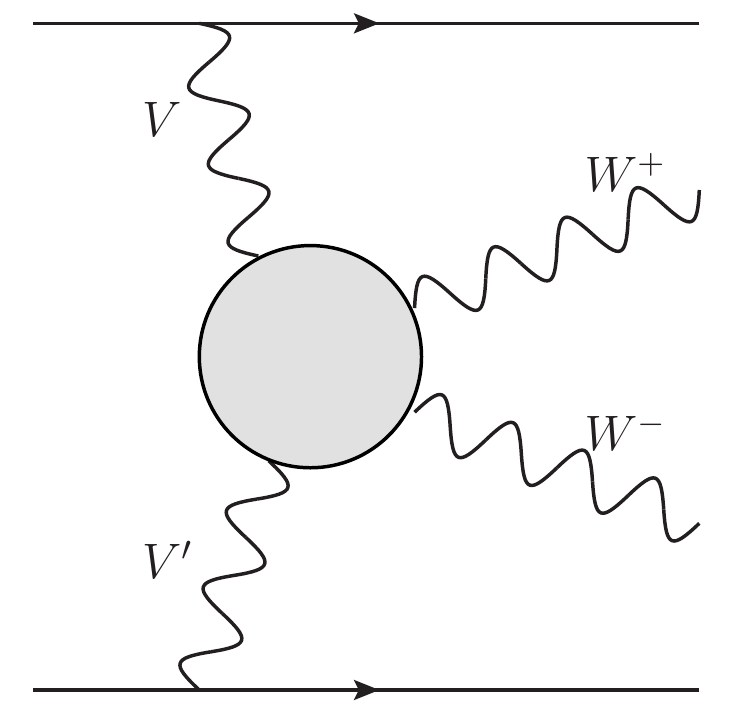}	
  	\caption{Diagrams depicting di-Higgs via VBF (left) and $W^+W^-$ VBS (right).}
  	\label{fig:diagrams}
  \end{figure}

A diagram depicting di-Higgs production via VBF $pp\to jjhh$
 is shown in \fig{fig:diagrams} (left). 
  We adopt the selection cuts listed in \tab{tab:cuts}.
  
 \begin{table}
\begin{tabular}{|l|c|c|c|c|} 
	\hline
$pp\to jjhh$ 		& $p_T(j) > 20\GeV$	& $|\eta(j)| < 5$	& $m(jj)>200\GeV$	& $\Delta R(jj)>0.4$	\\
					& $p_T(h)>30\GeV$	& $|\eta(h)|<3.5$ 	& 					&	\\
	\hline
$pp\to jjW^+W^-$ 	& $p_T(j) > 20\GeV$	& $|\eta(j)| < 5$	& $m(jj)>250\GeV$	& $\Delta\eta(jj)>2.5$	\\	
					& $p_T(w)>30\GeV$ 	& $|\eta(W^\pm)|<3.5$	& $m(W^+W^-)>500\GeV$ & \\
	\hline
$pp\to jj\eta\eta$ 	& $p_T(j) > 20\GeV$	& $|\eta(j)| < 5$	& $m(jj)>200\GeV$	& $\Delta R(jj)>0.4$	\\	
					& $p_T(\eta)>30\GeV$ 	& $|\eta(\eta)|<3.5$	& 		&  \\
	\hline
$pp\to jj\eta_5^{++}\eta_5^{--}$ 	& $p_T(j) > 20\GeV$	& $|\eta(j)| < 5$	& $m(jj)>200\GeV$	& $\Delta R(jj)>0.4$	\\	
	\hline
\end{tabular}
\caption{Selection cuts.}
\label{tab:cuts}
 \end{table}

\begin{figure}[htbp]
\centering
\includegraphics[width=0.32\textwidth]{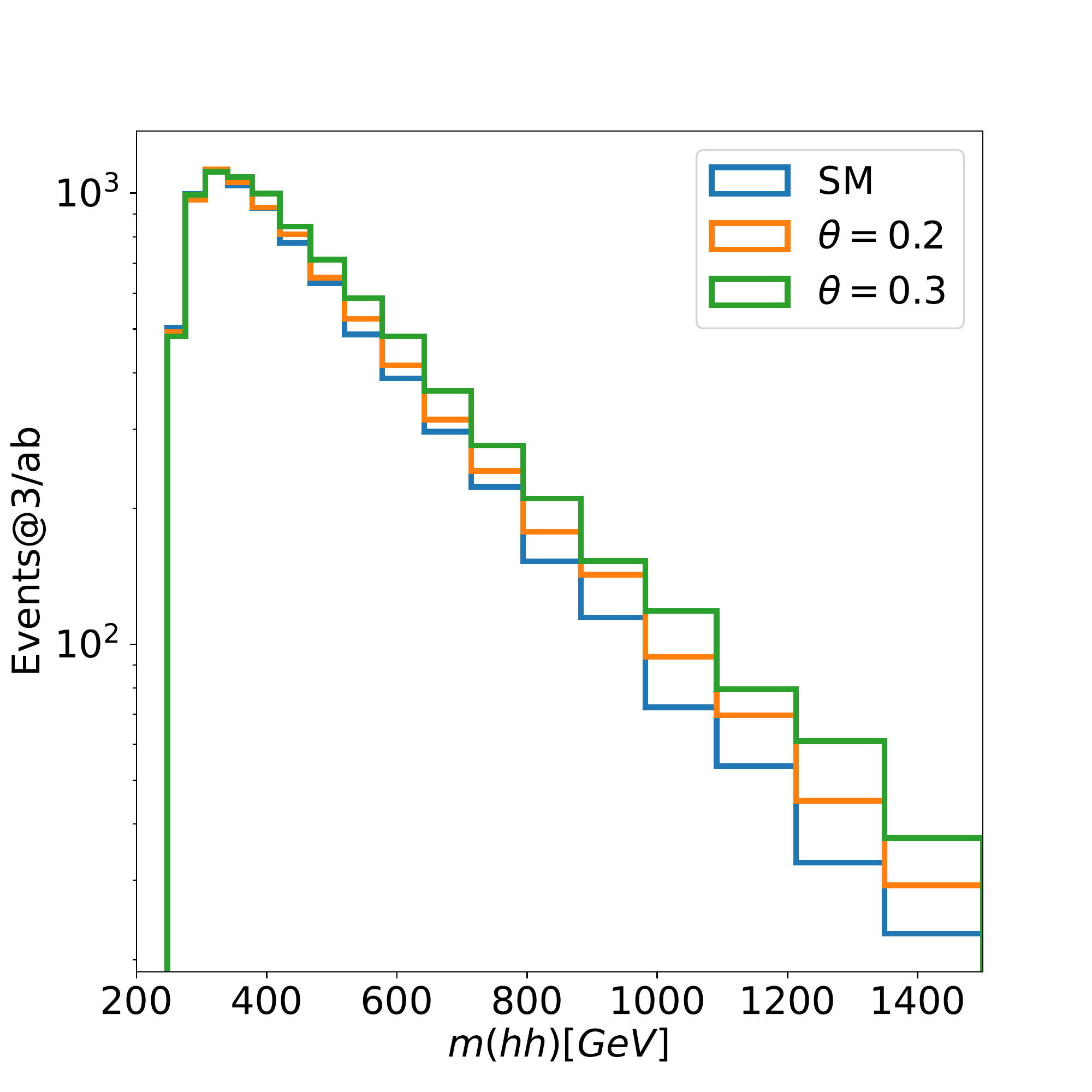}
\includegraphics[width=0.32\textwidth]{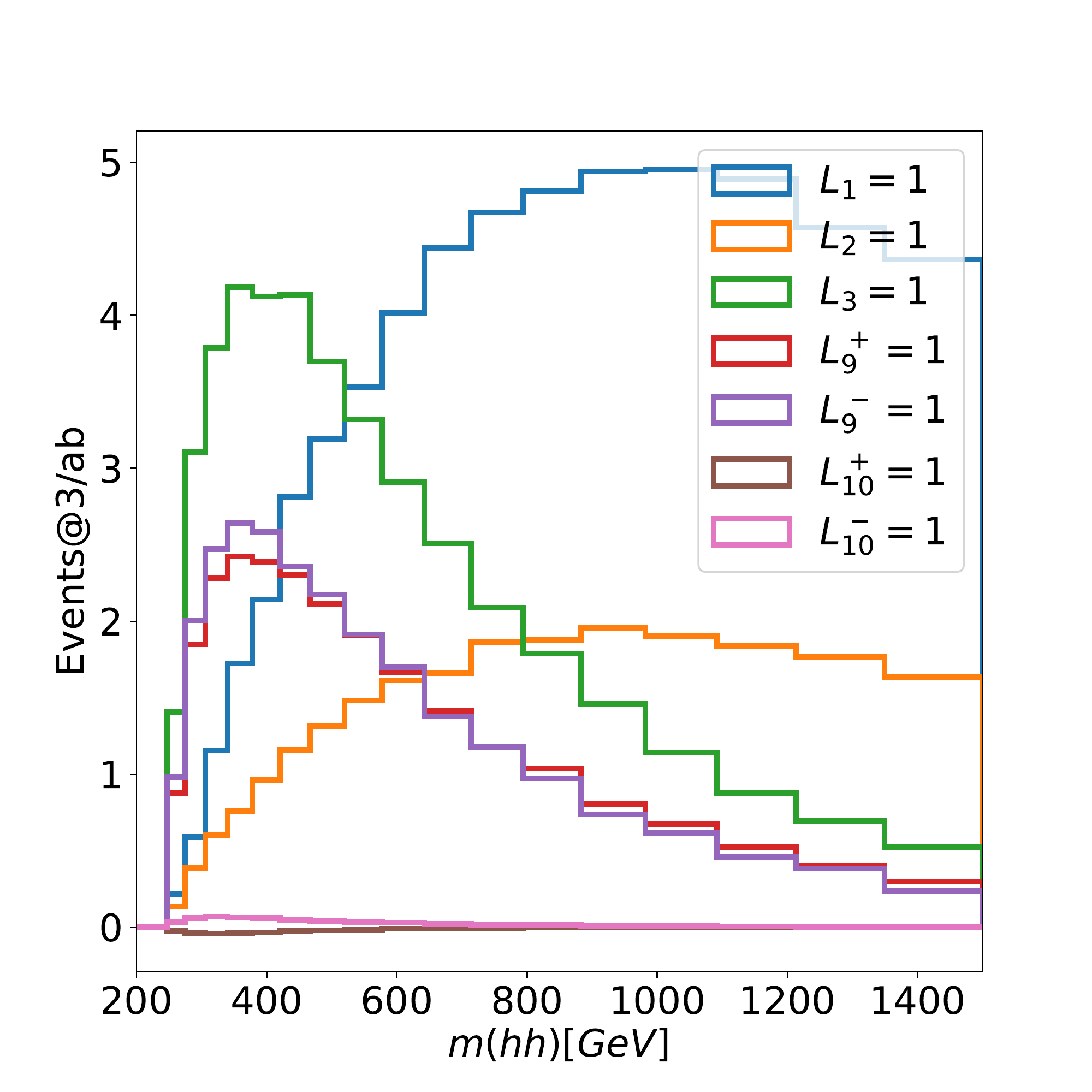}
\includegraphics[width=0.32\textwidth]{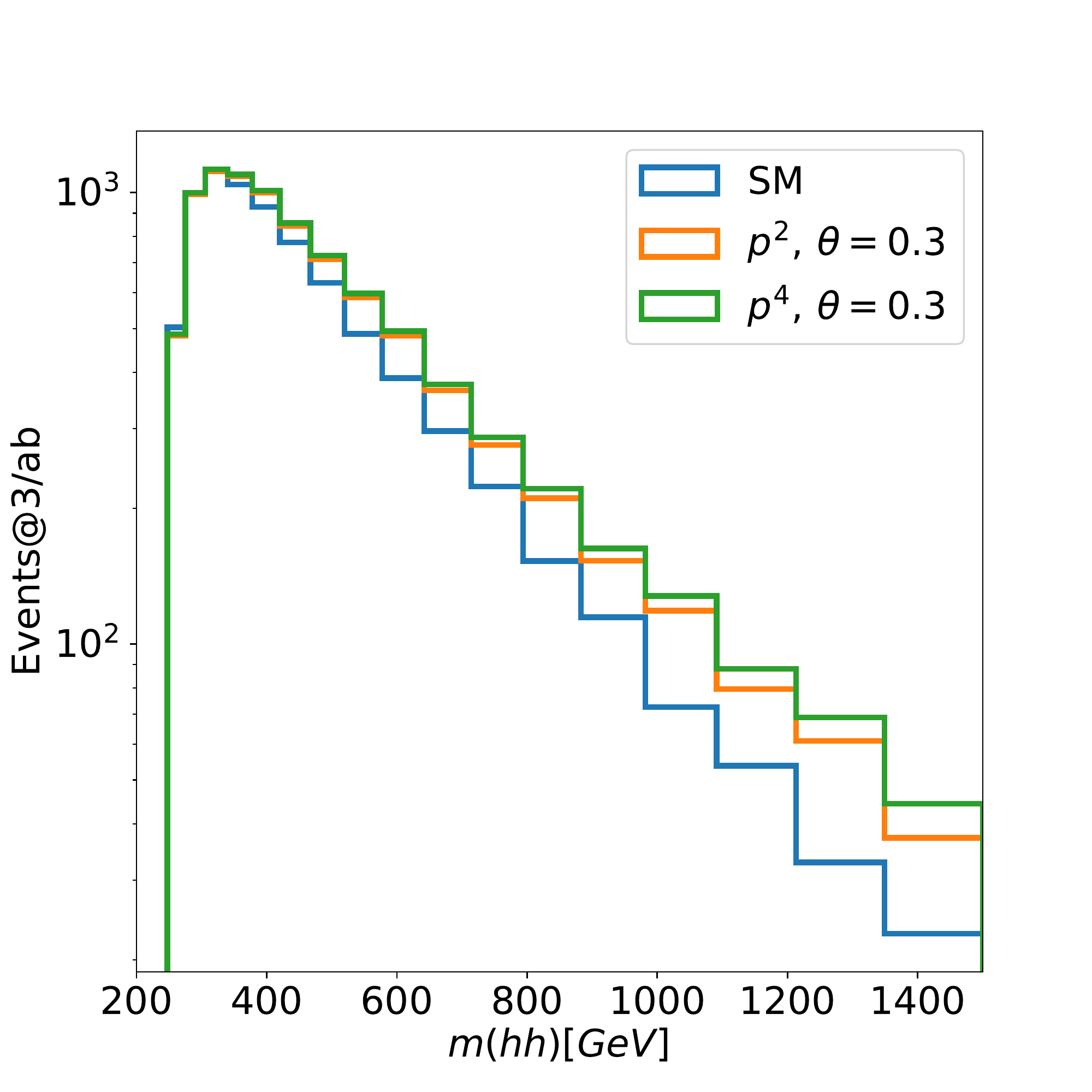}

\includegraphics[width=0.32\textwidth]{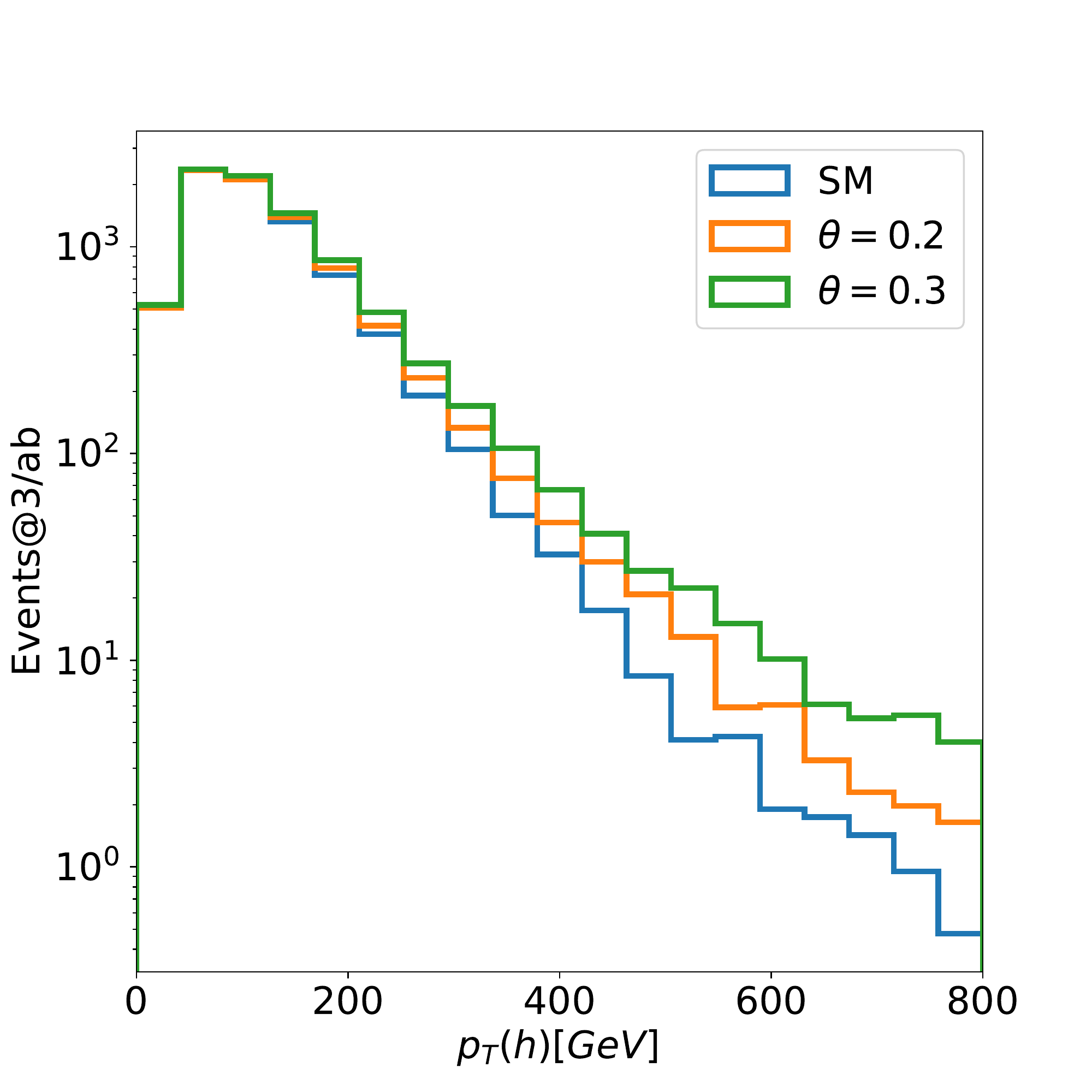}
\includegraphics[width=0.32\textwidth]{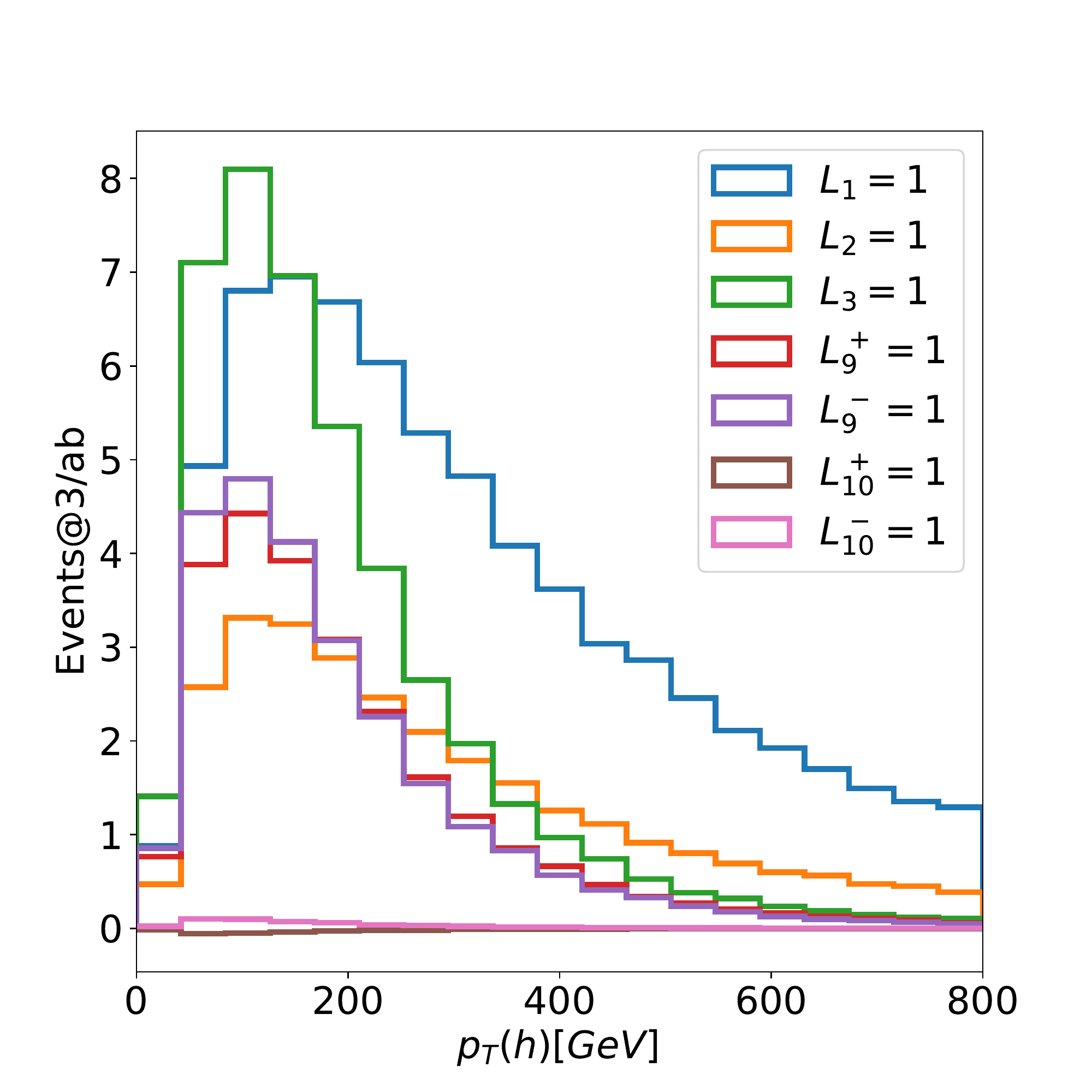}
\includegraphics[width=0.32\textwidth]{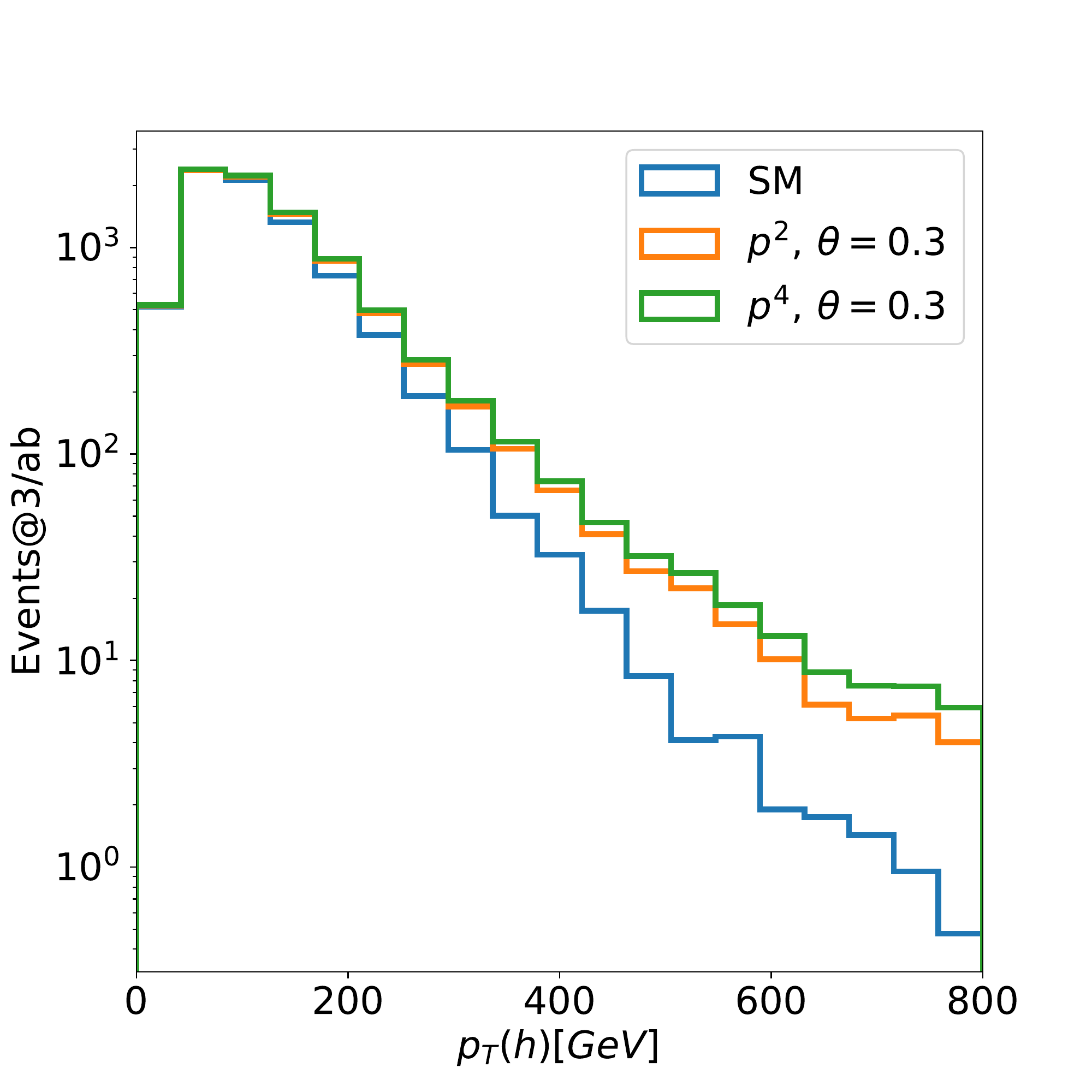}

\includegraphics[width=0.32\textwidth]{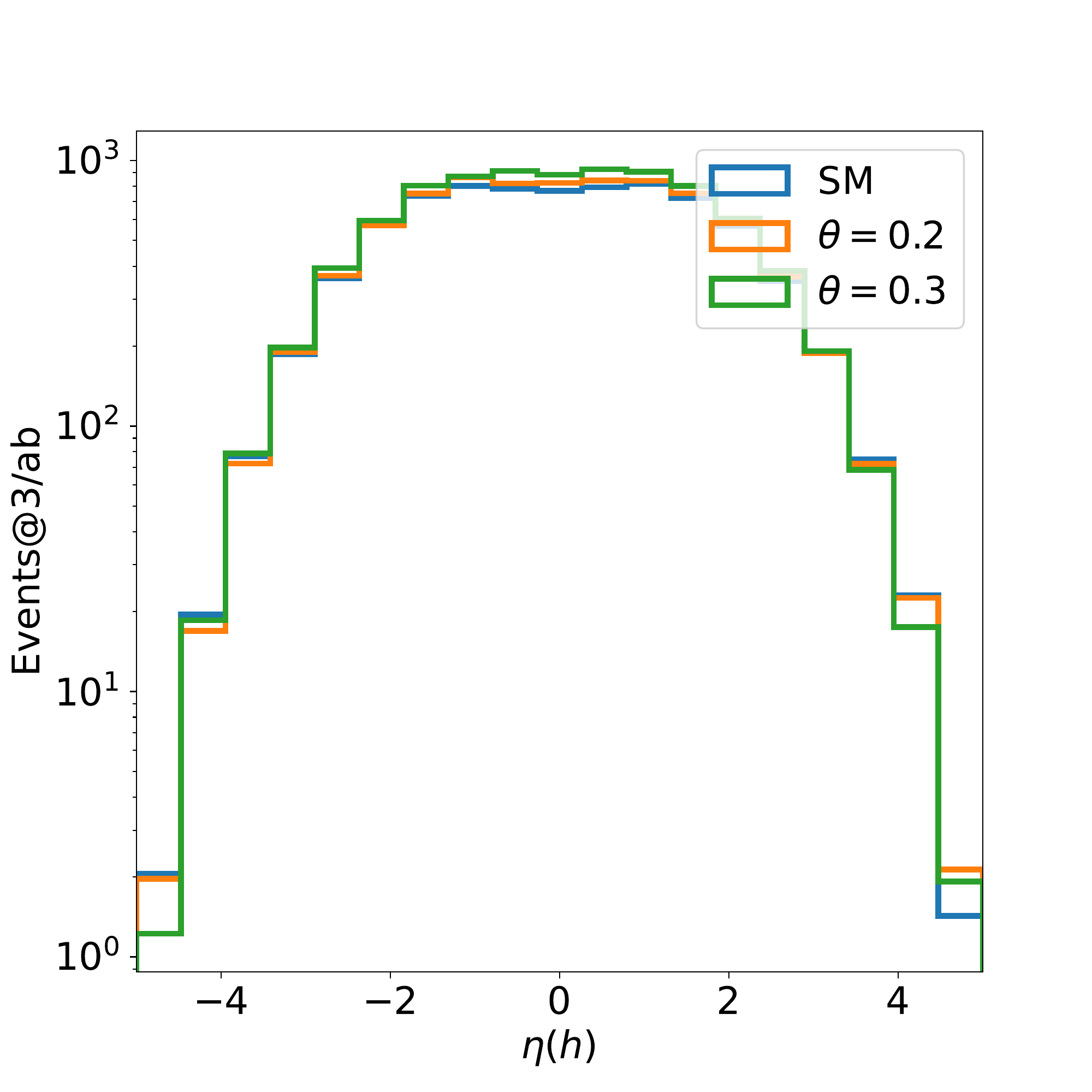}
\includegraphics[width=0.32\textwidth]{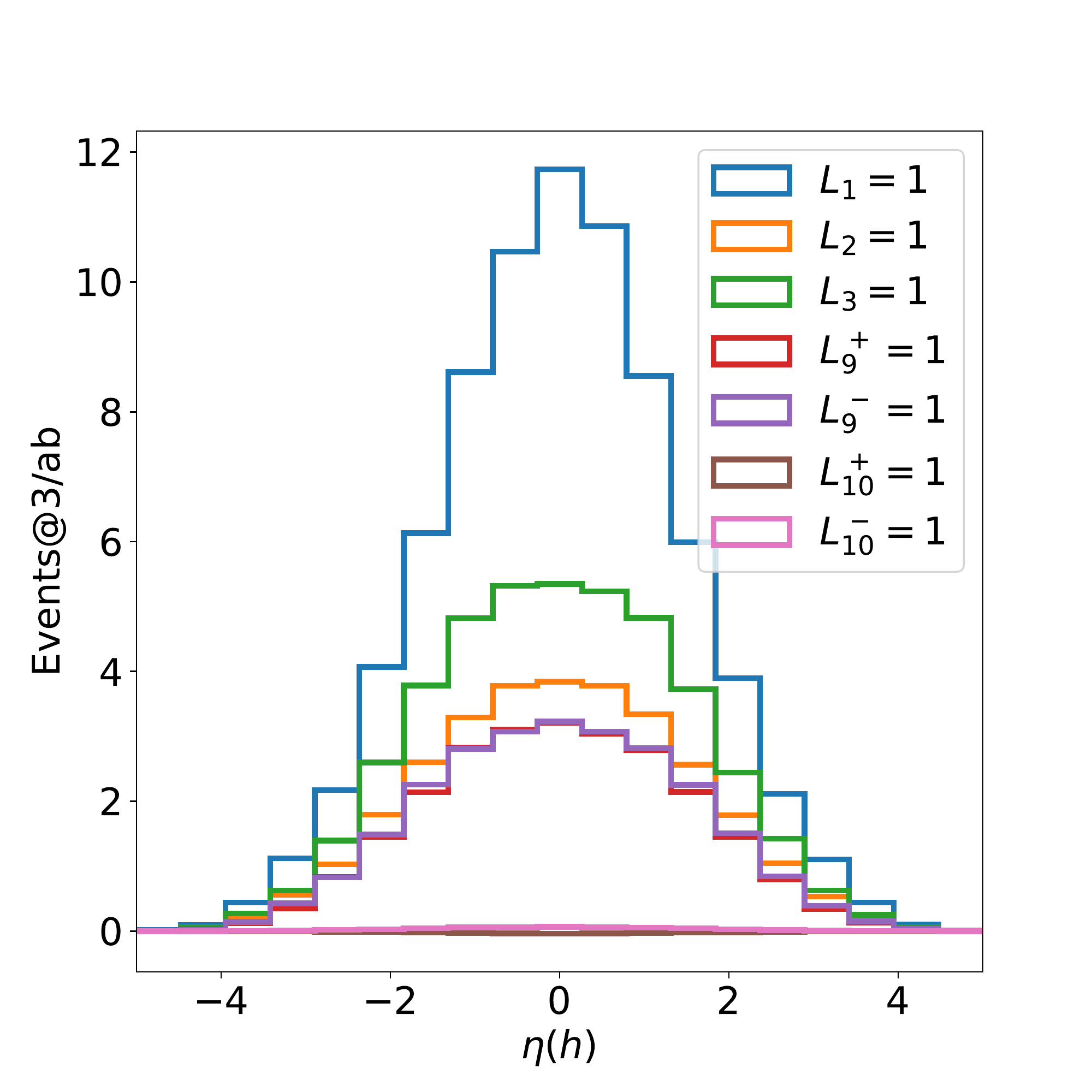}
\includegraphics[width=0.32\textwidth]{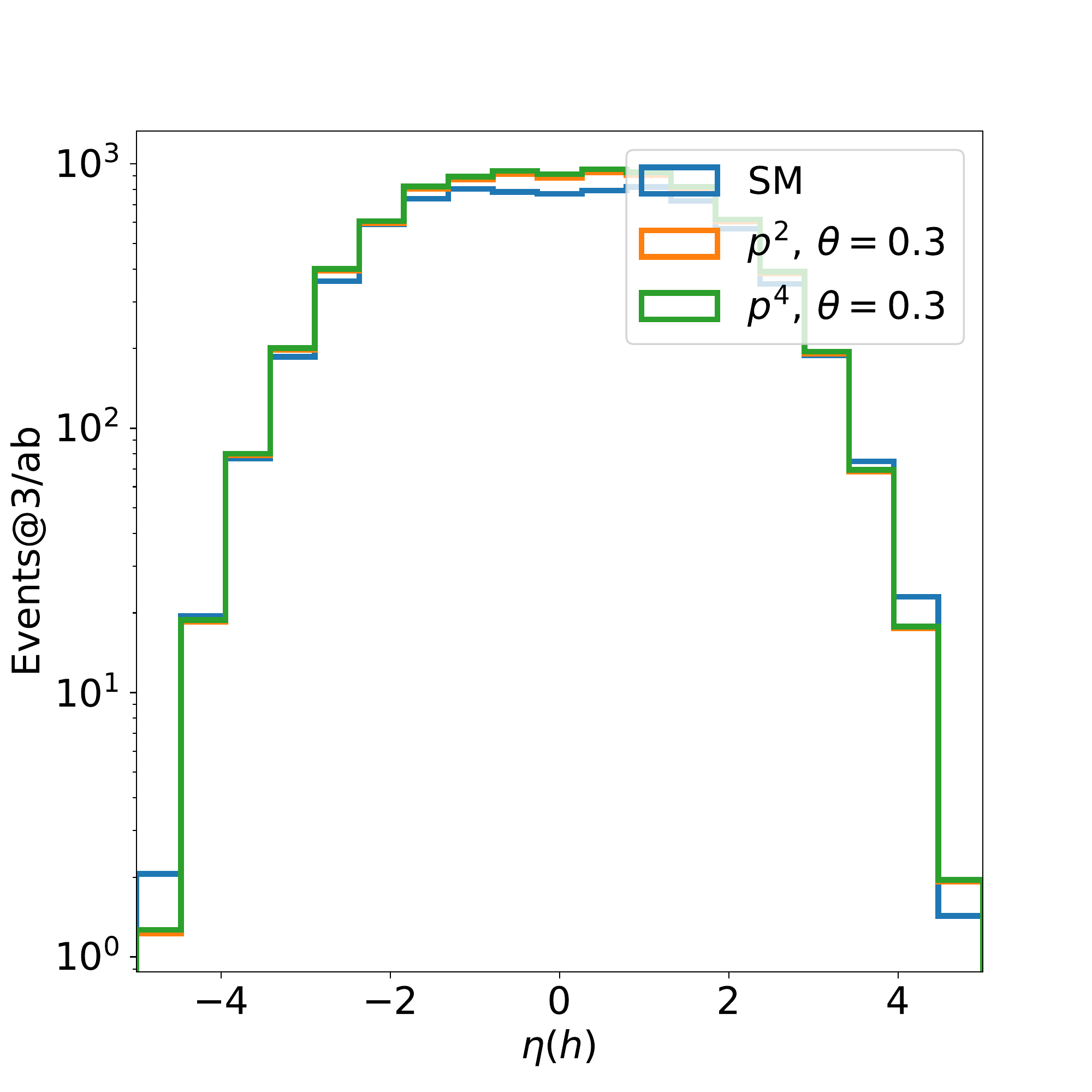}

\caption{Distributions of di-Higgs via VBF.}
\label{fig:dihiggs}
\end{figure}

Predictions for the distributions of the invariant mass of the di-Higgs system $m(hh)$, the transverse momentum $p_T(h)$, and the pseudo-rapidity $\eta(h)$ of the hardest Higgs boson are shown in \fig{fig:dihiggs}, on the top, center and bottom rows respectively. 
The left column compare the predictions of the LO Lagrangian for different values of $\theta=0.2,\,0.3$ and the SM. The usual growing behavior with energy is observed in the $m(hh)$ distribution for the CH scenarios. 
In the middle column the contribution of each of the NLO operators to each observable are shown separately for each  $L_i$ coefficient set to 1. These should be summed linearly to the distributions. 
It is interesting to notice that the genuinely quartic couplings give the largest contributions at high energy, but they are suppressed near threshold production. %
In the right column the full NLO prediction with all $L_i=1$ are compared to the LO predictions.

\begin{figure}[htbp]
\centering
\includegraphics[width=0.32\textwidth]{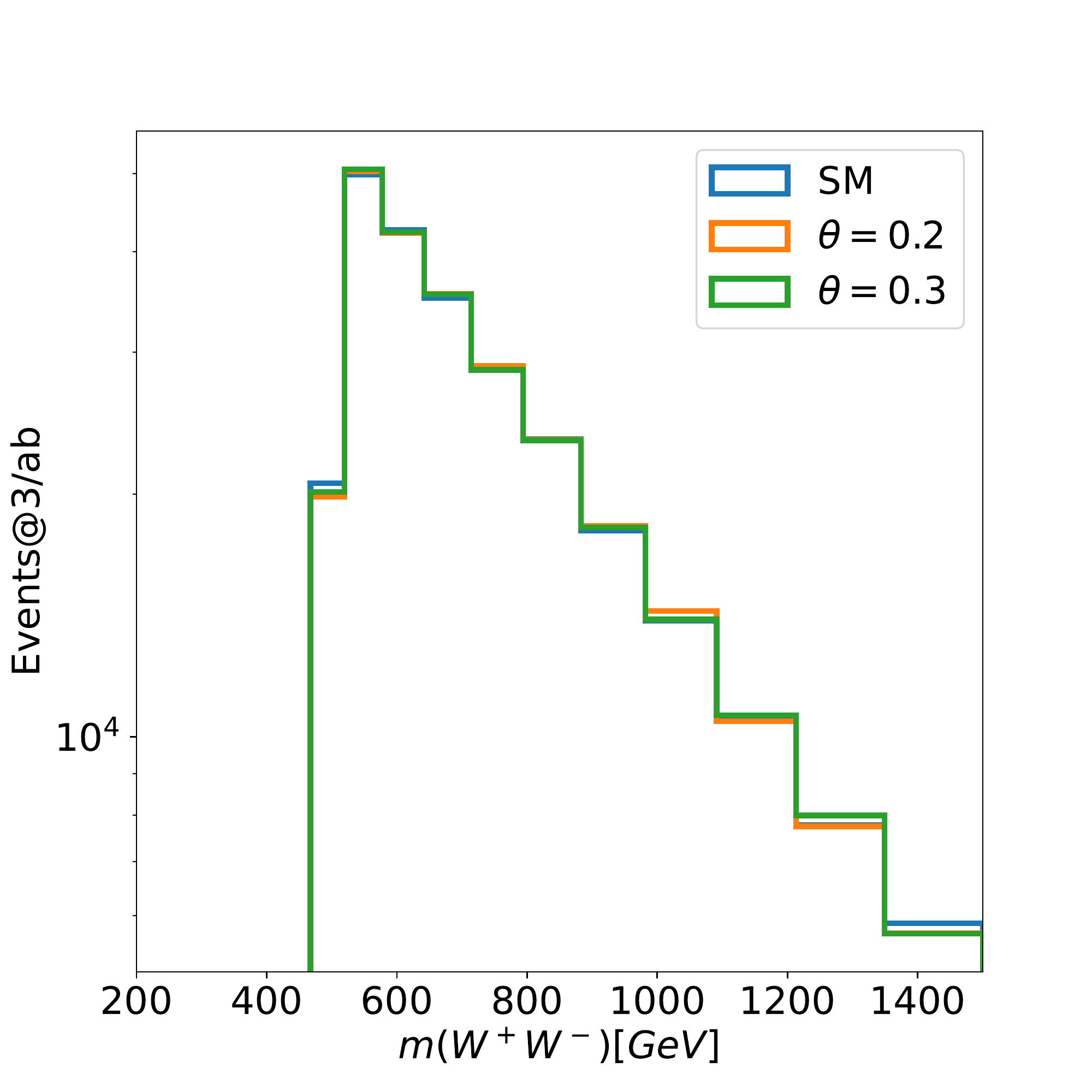}
\includegraphics[width=0.32\textwidth]{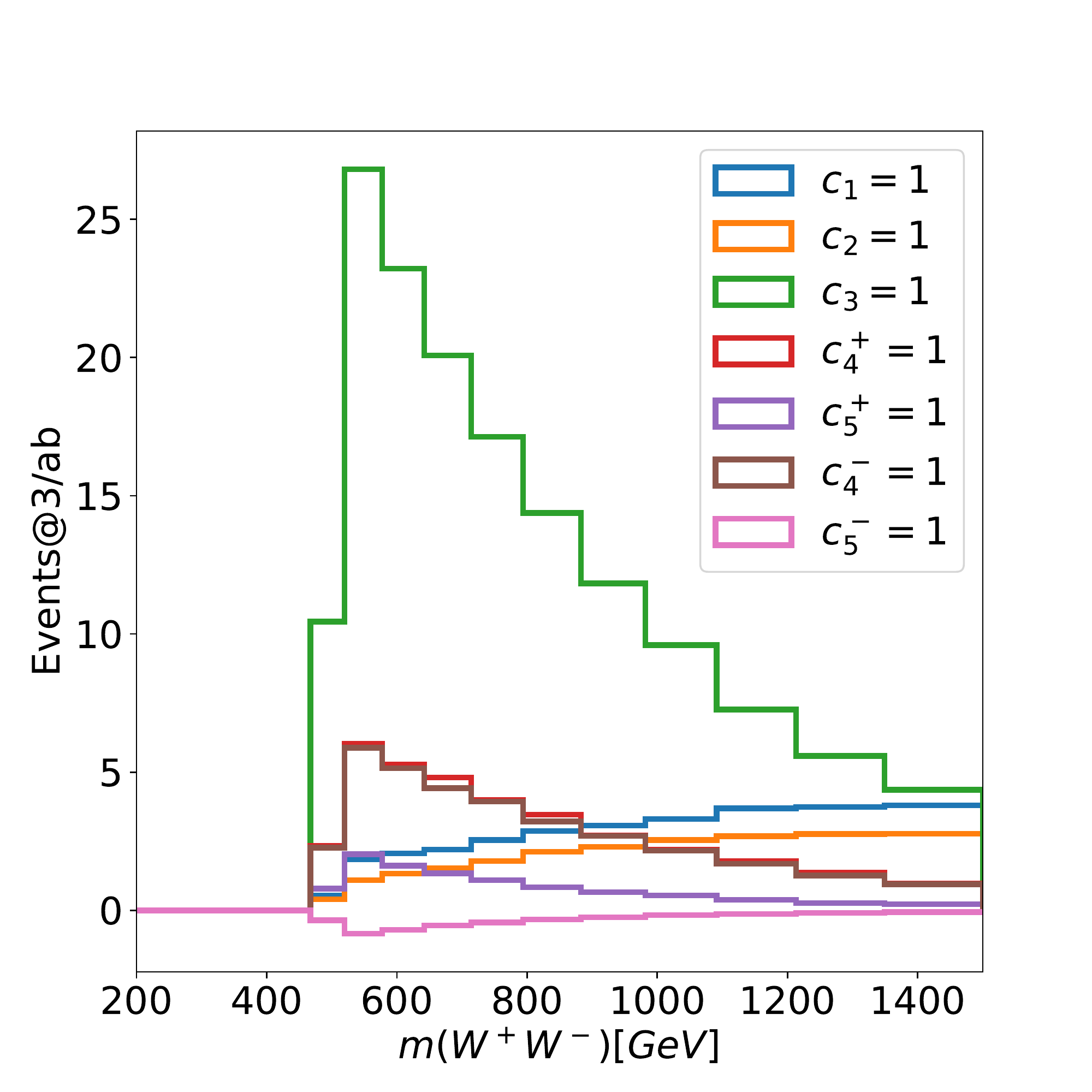}
\includegraphics[width=0.32\textwidth]{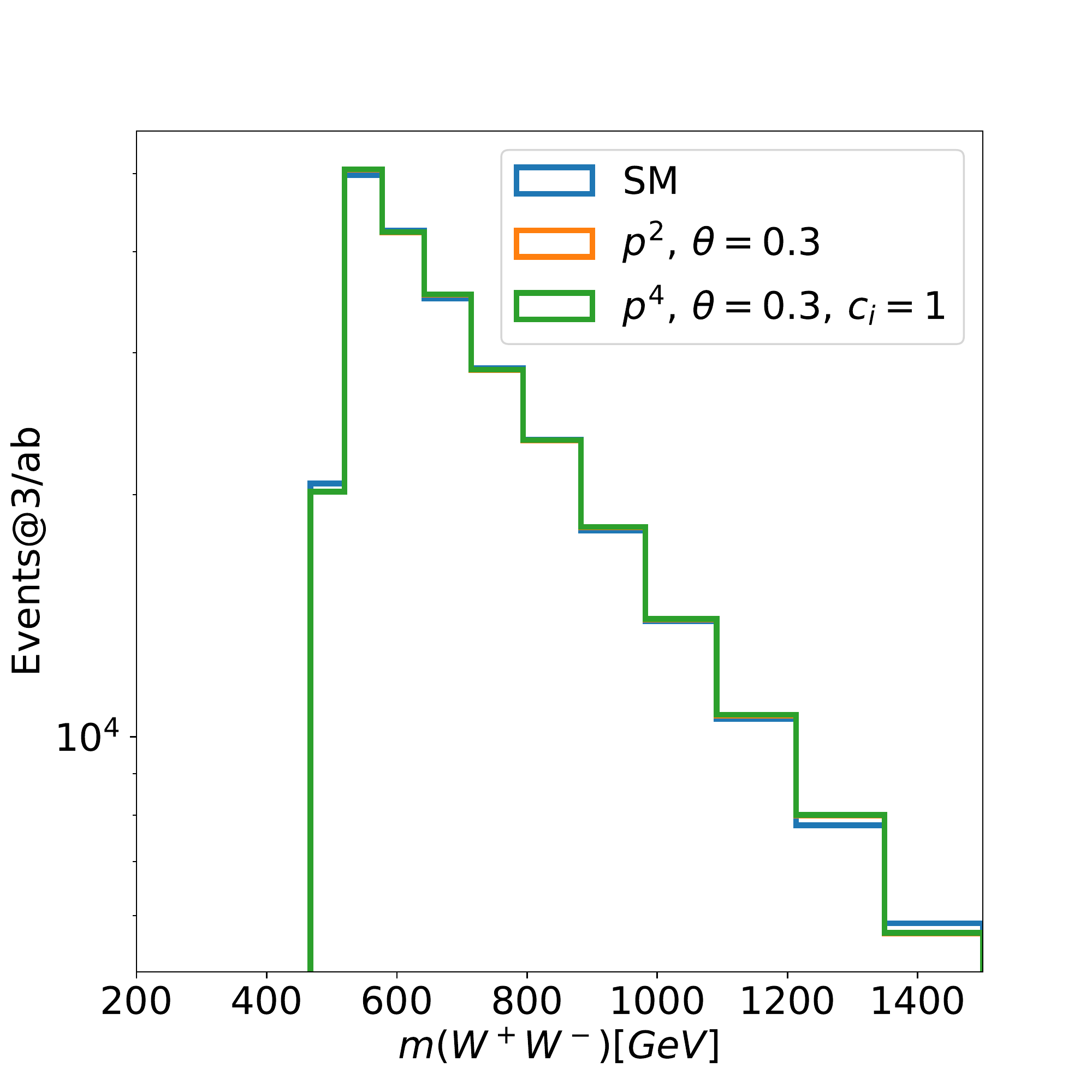}

\includegraphics[width=0.32\textwidth]{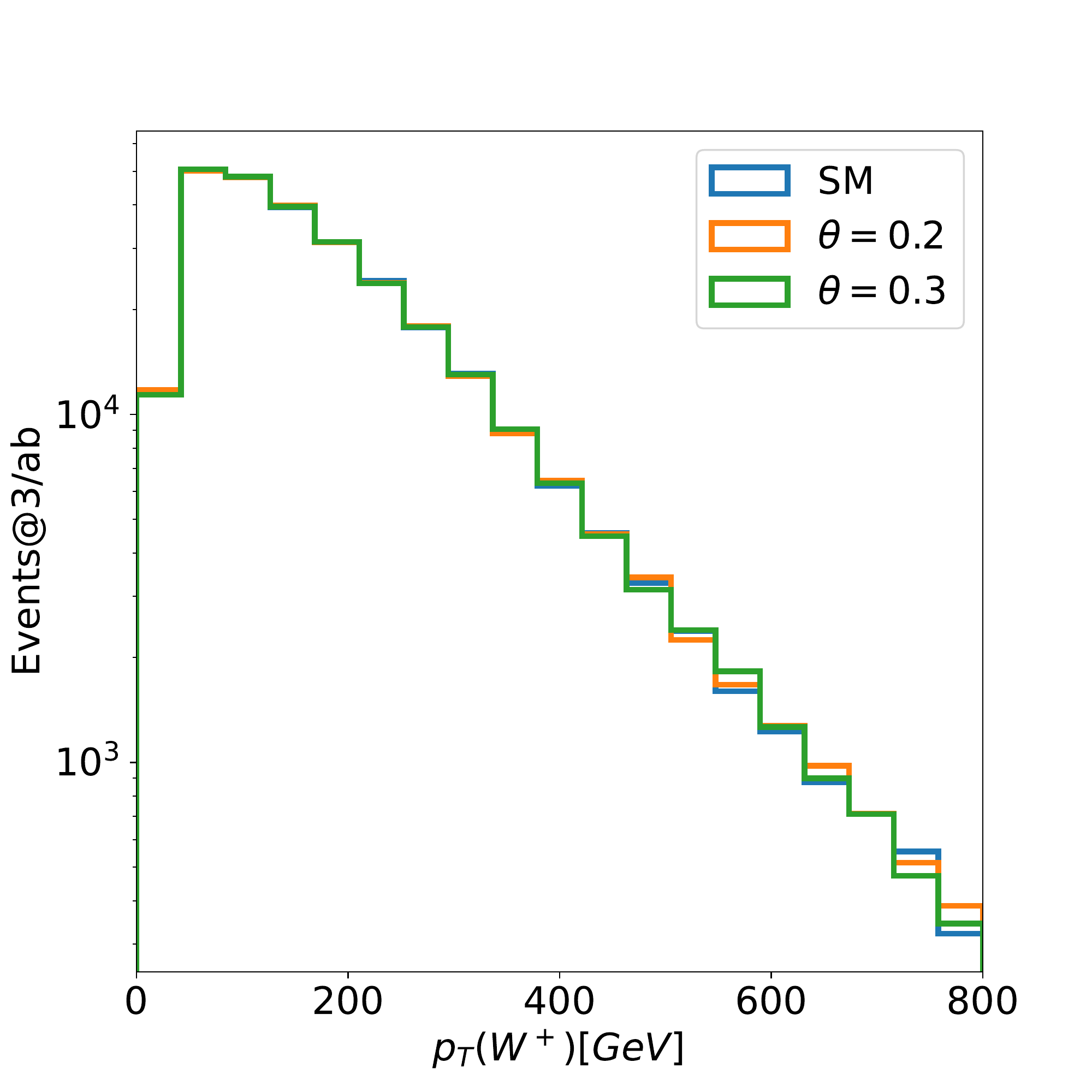}
\includegraphics[width=0.32\textwidth]{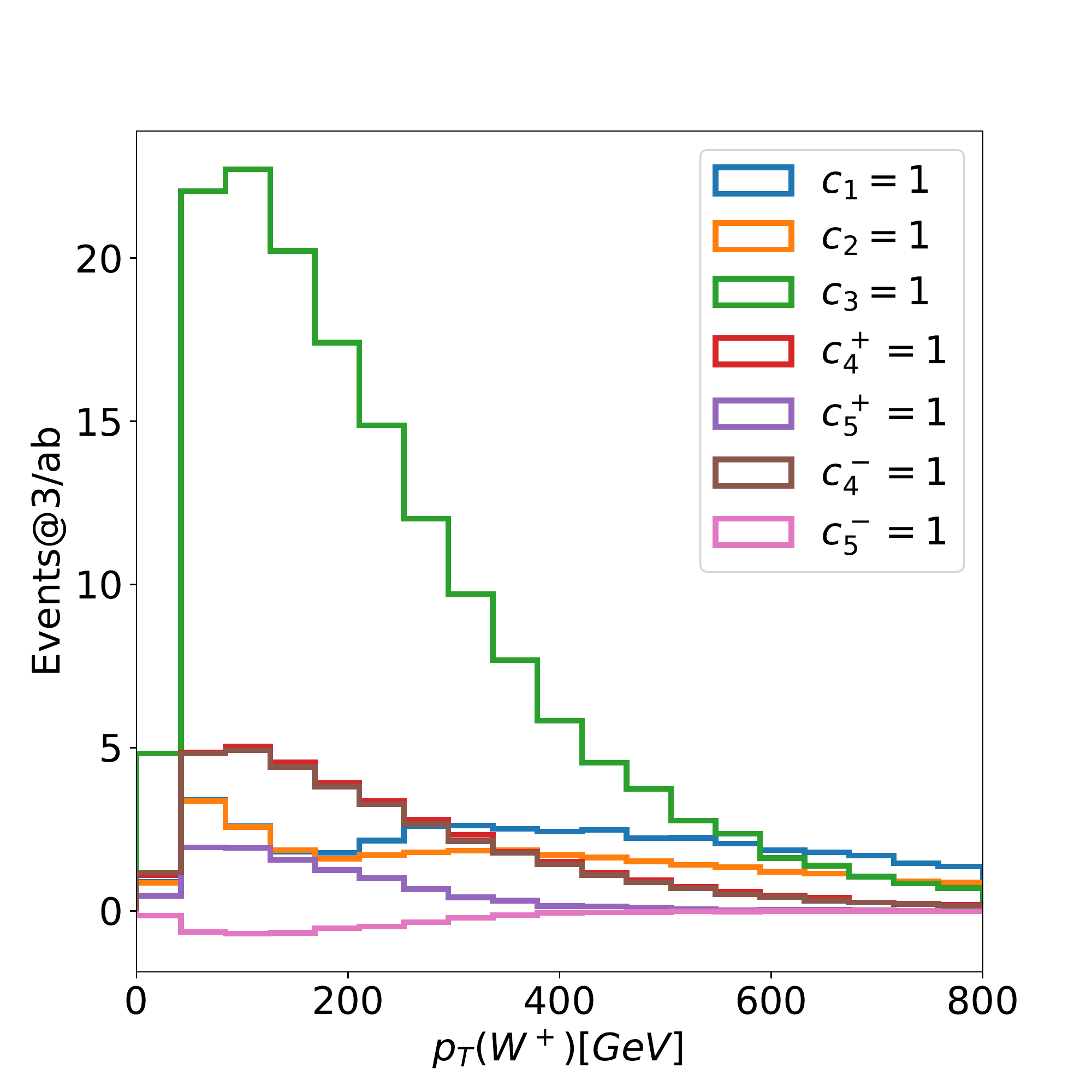}
\includegraphics[width=0.32\textwidth]{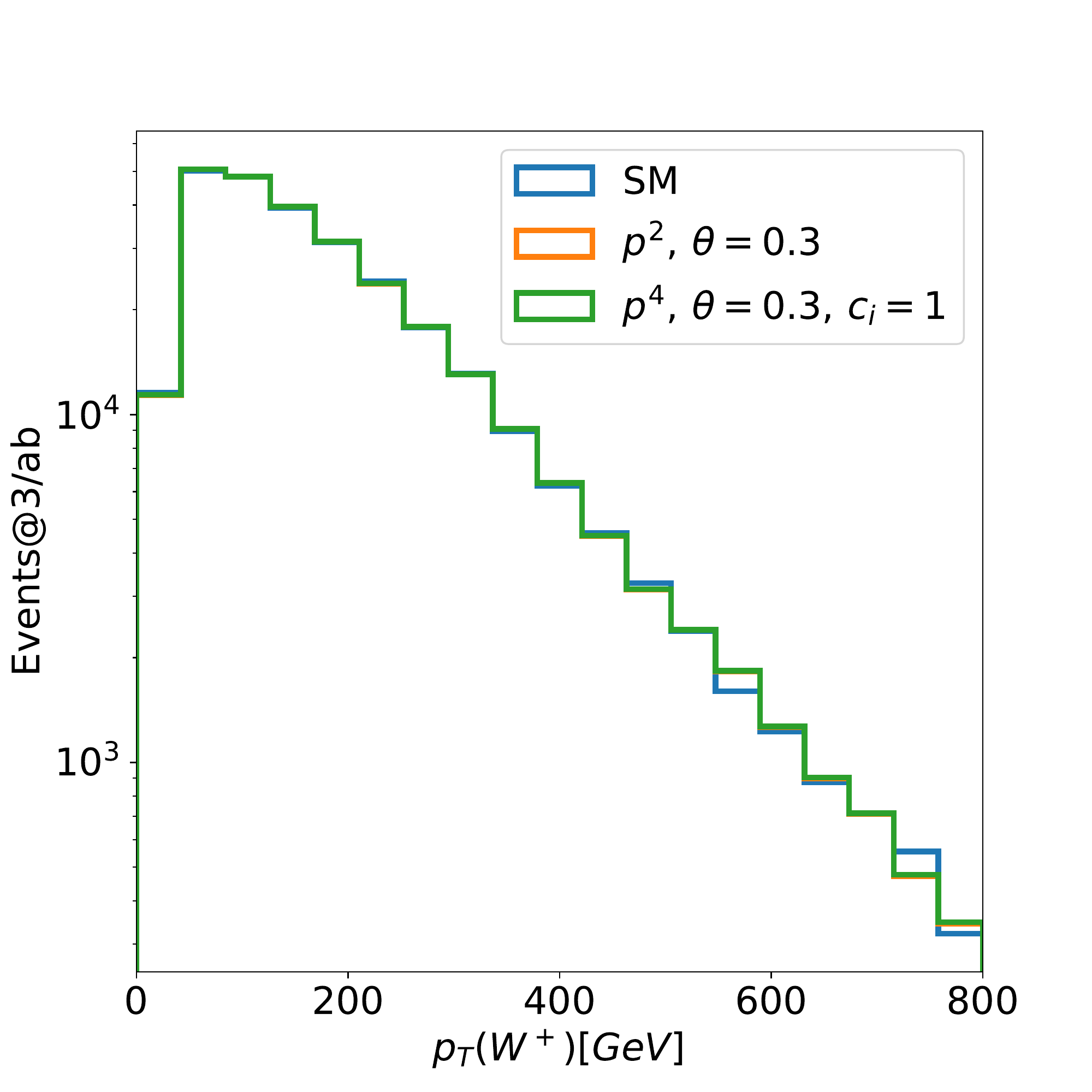}

\includegraphics[width=0.32\textwidth]{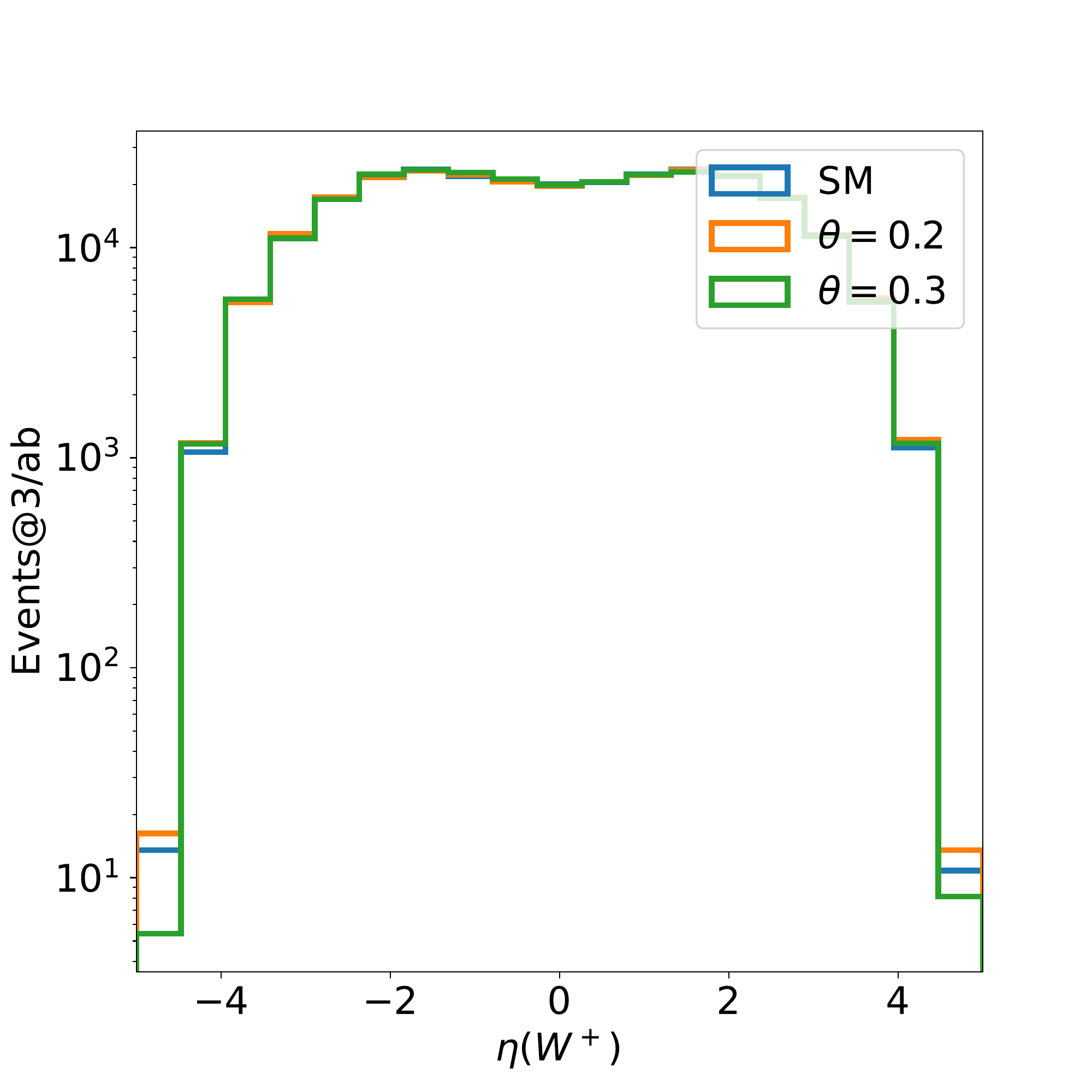}
\includegraphics[width=0.32\textwidth]{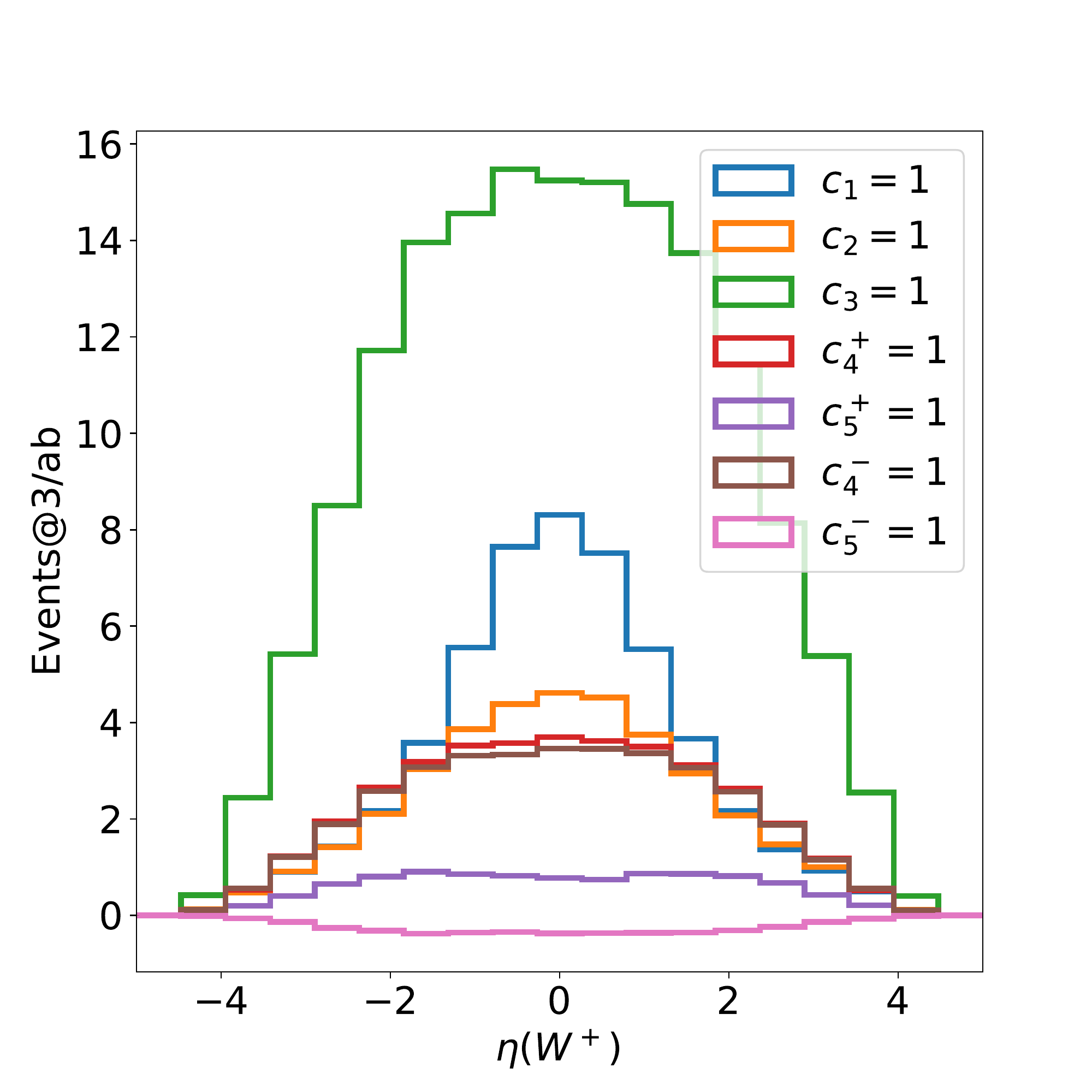}
\includegraphics[width=0.32\textwidth]{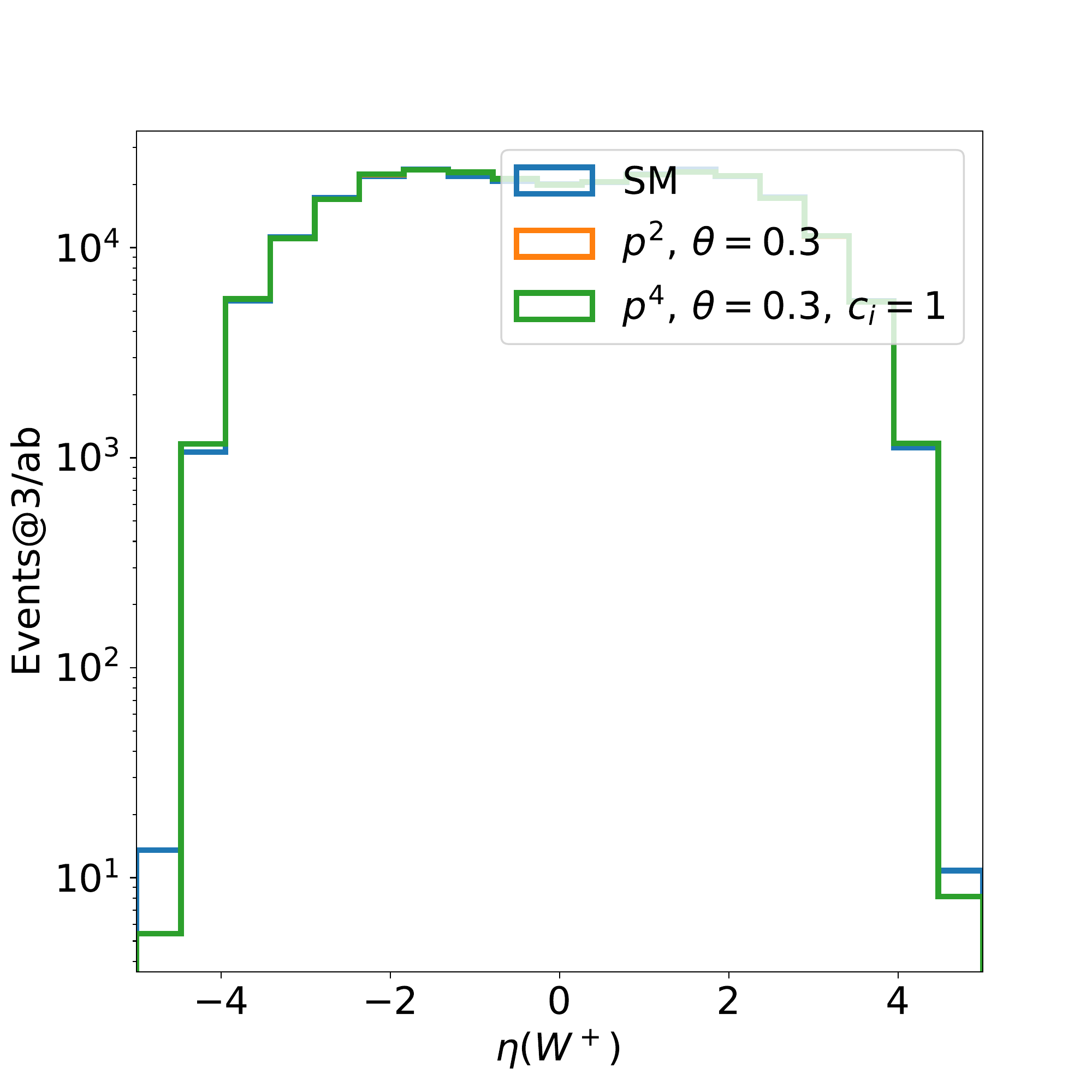}

\caption{Unpolarized VBS.}
\label{fig:vbs}
\end{figure}

\begin{figure}[htbp]
	\centering
	\includegraphics[width=0.32\textwidth]{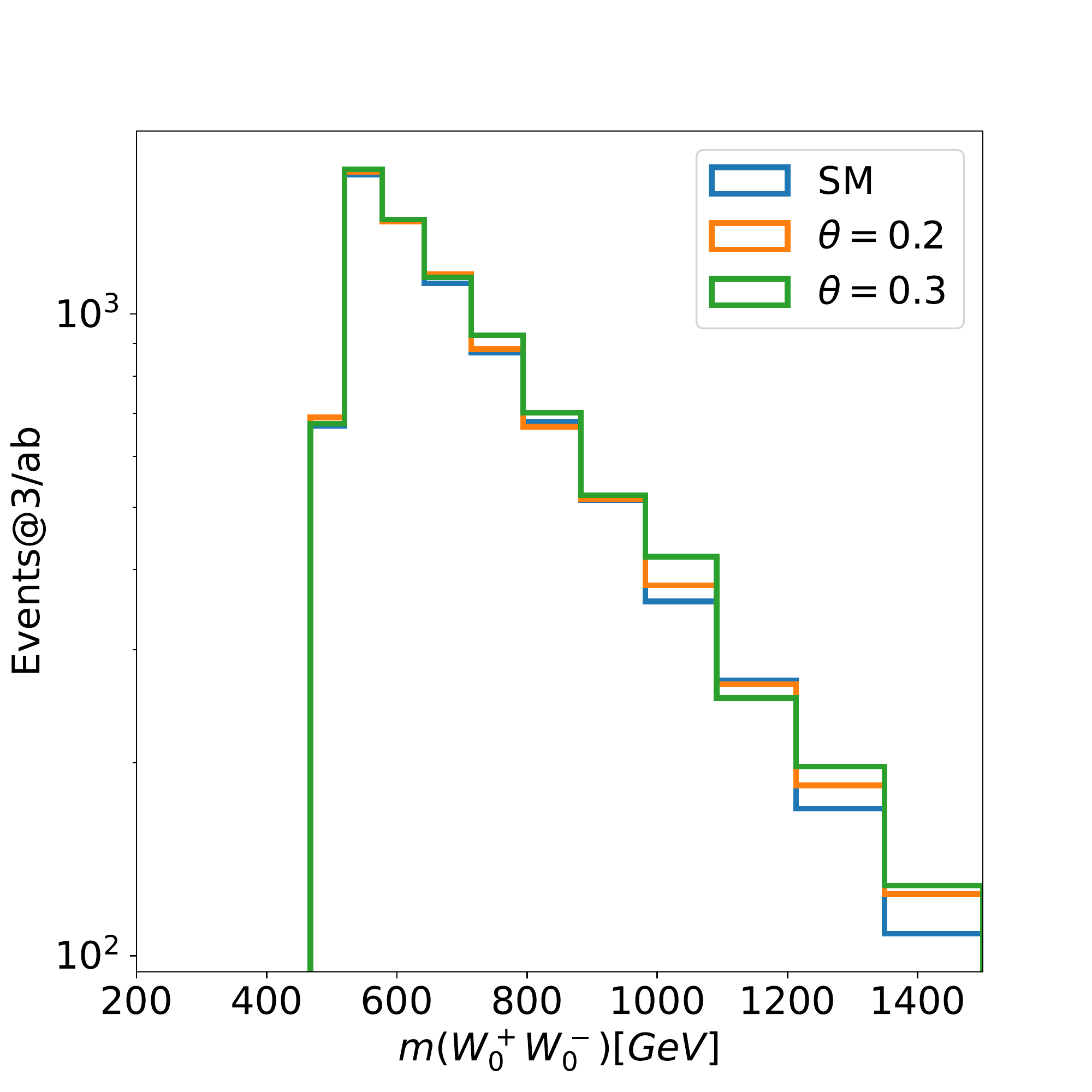}
	\includegraphics[width=0.32\textwidth]{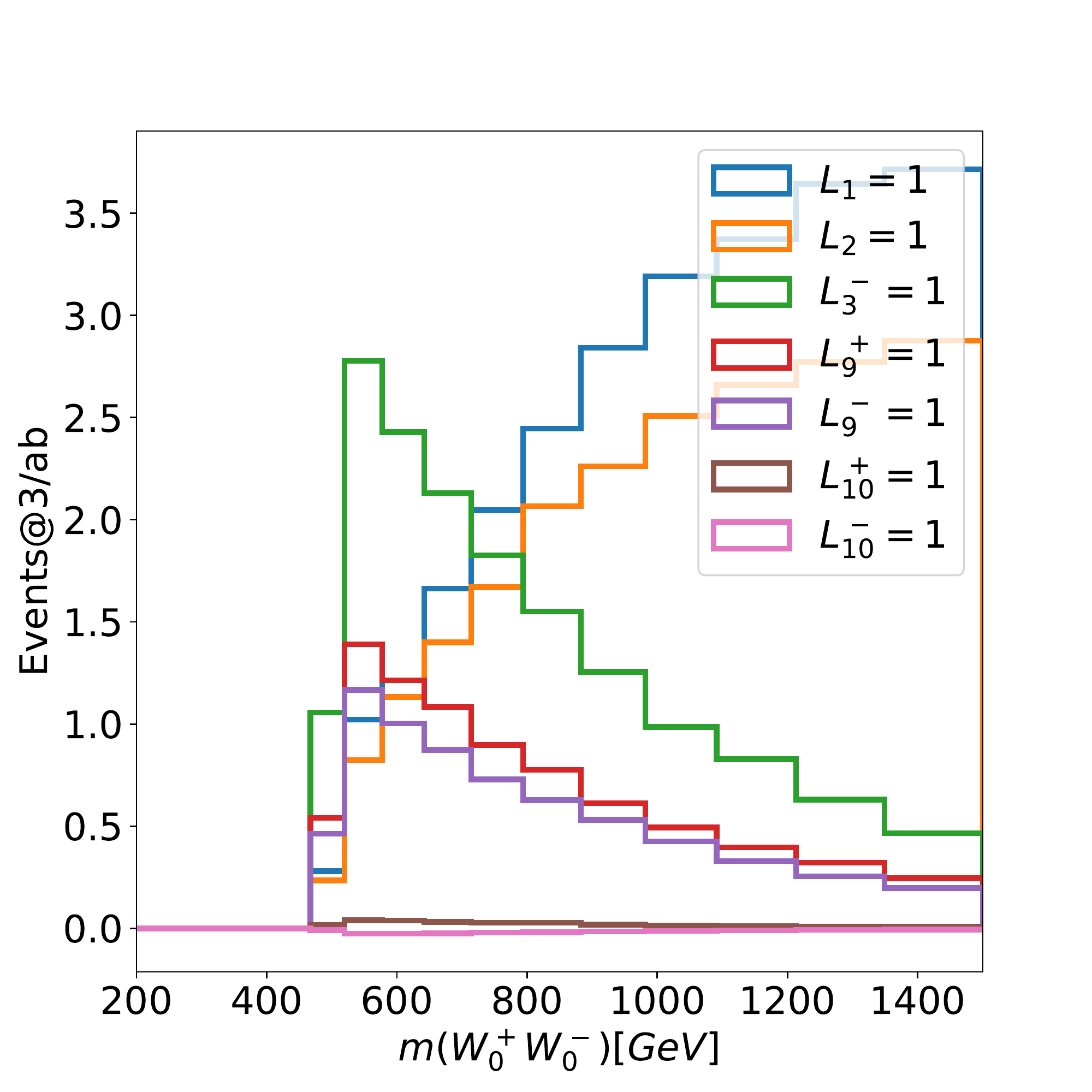}
	\includegraphics[width=0.32\textwidth]{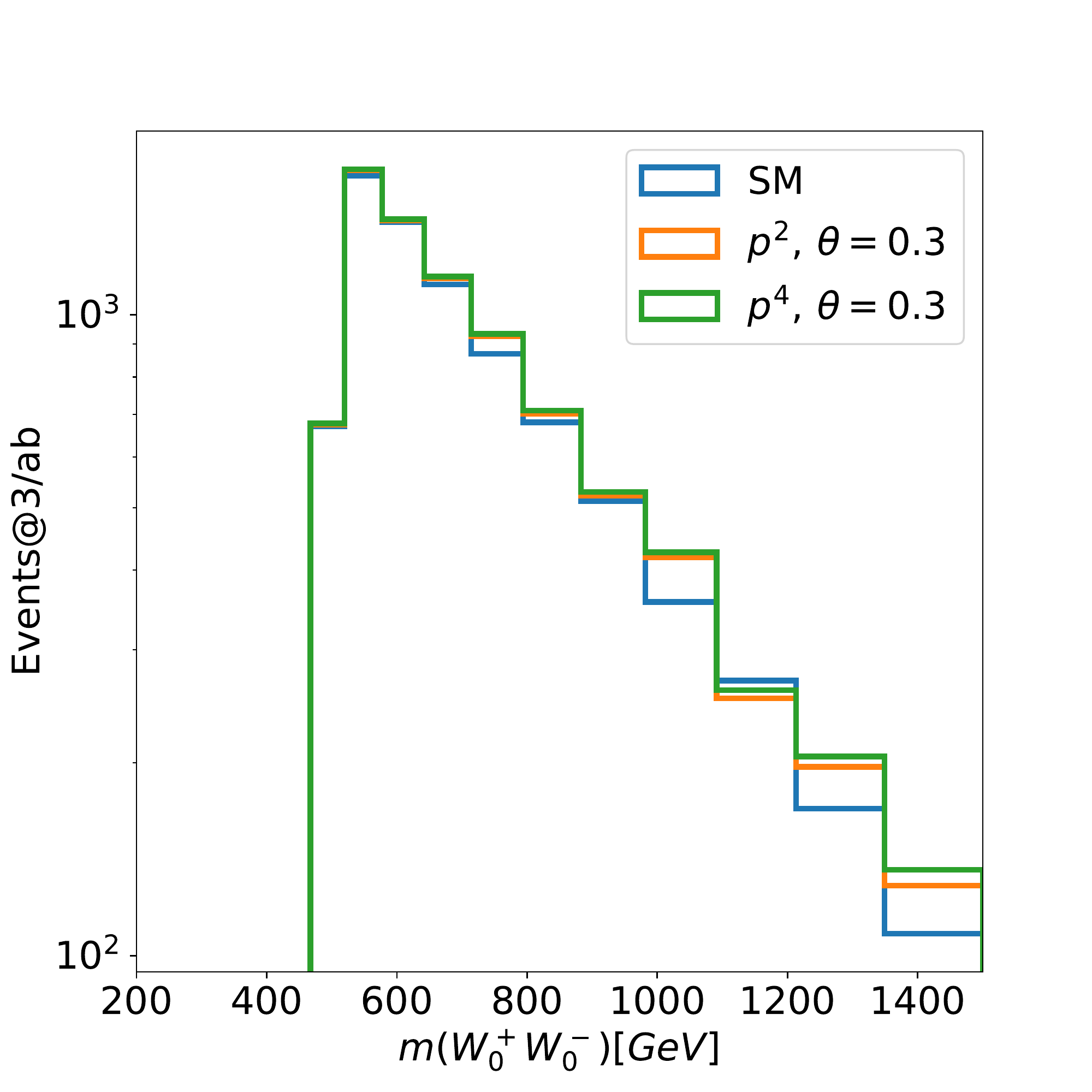}
	
	\includegraphics[width=0.32\textwidth]{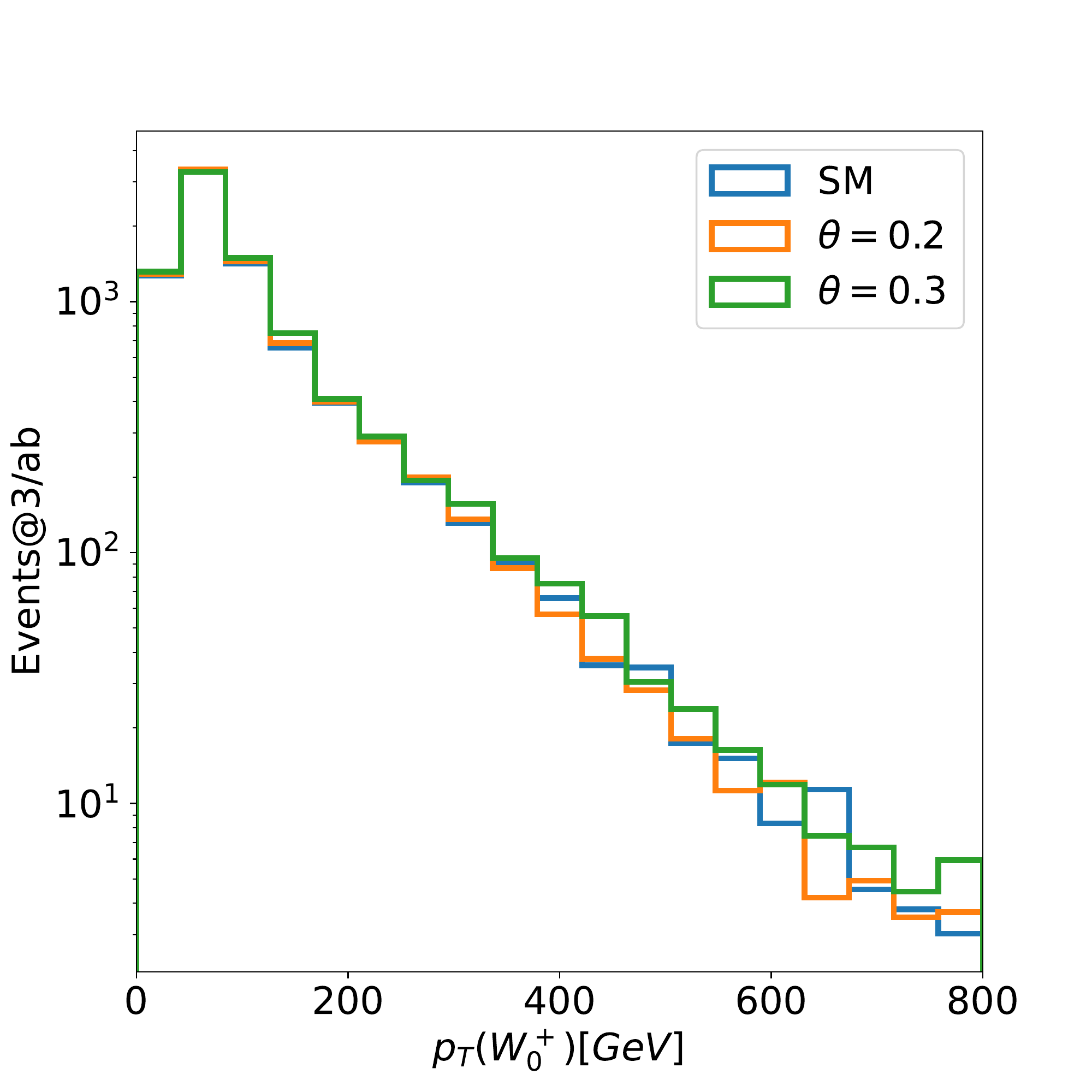}
	\includegraphics[width=0.32\textwidth]{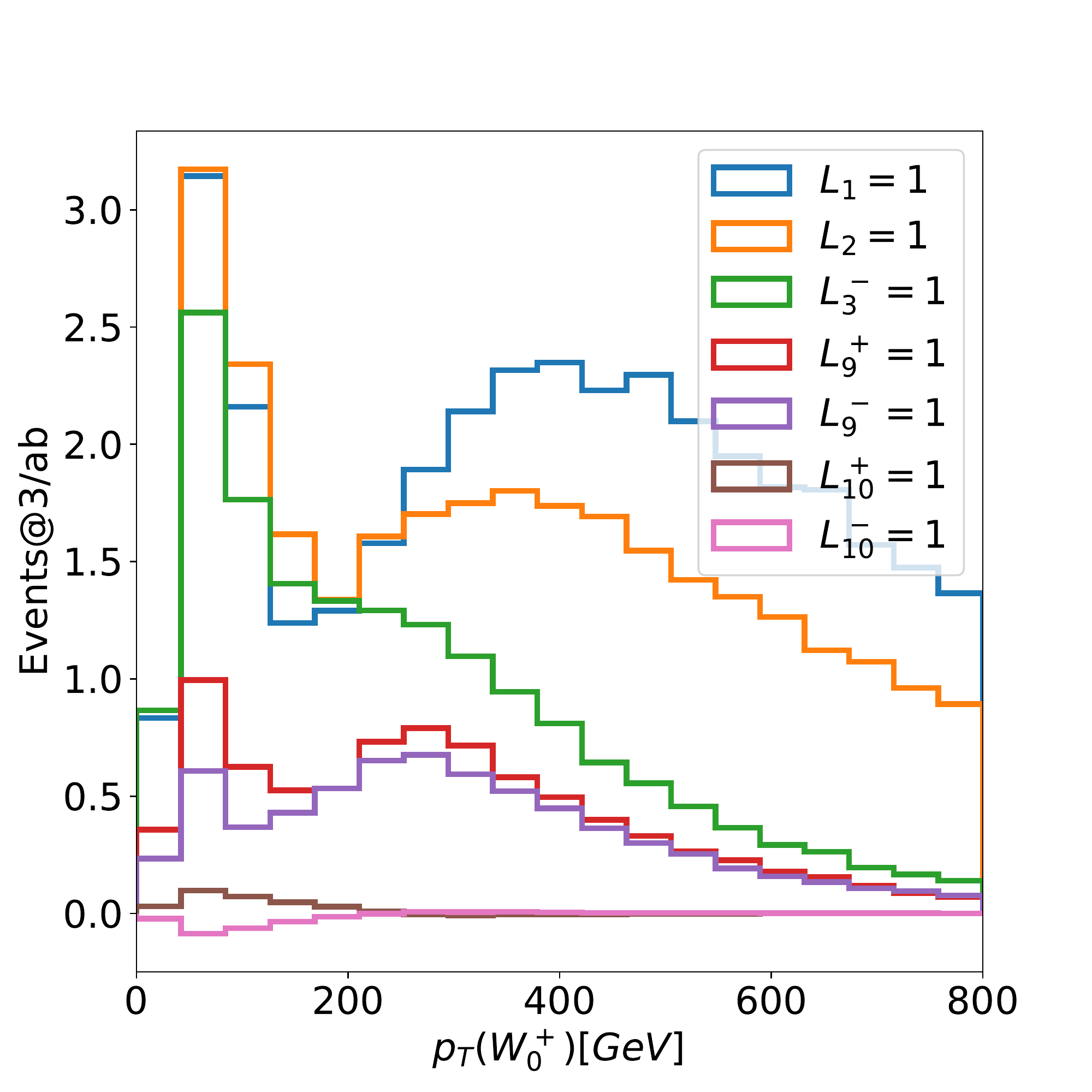}
	\includegraphics[width=0.32\textwidth]{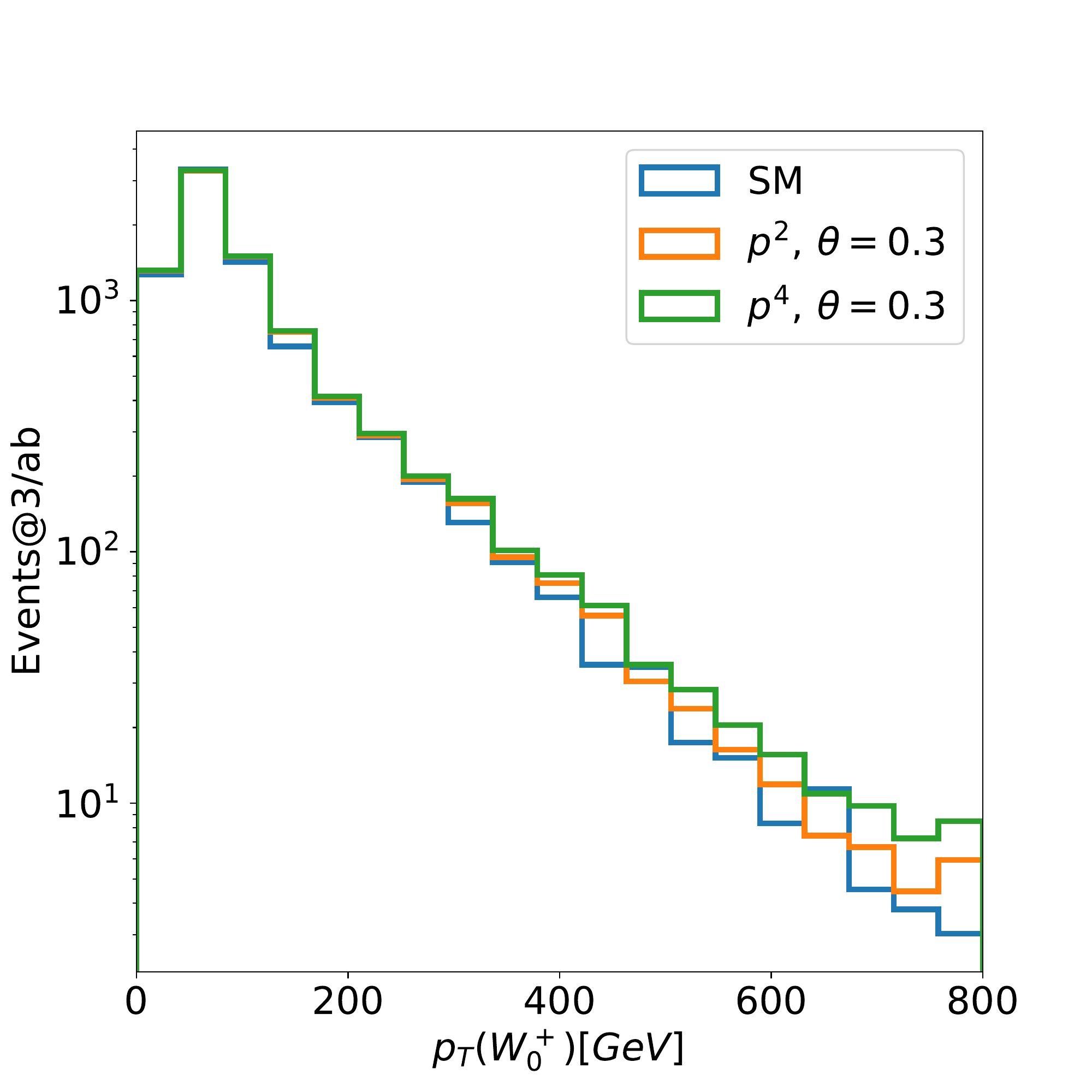}
	
	\includegraphics[width=0.32\textwidth]{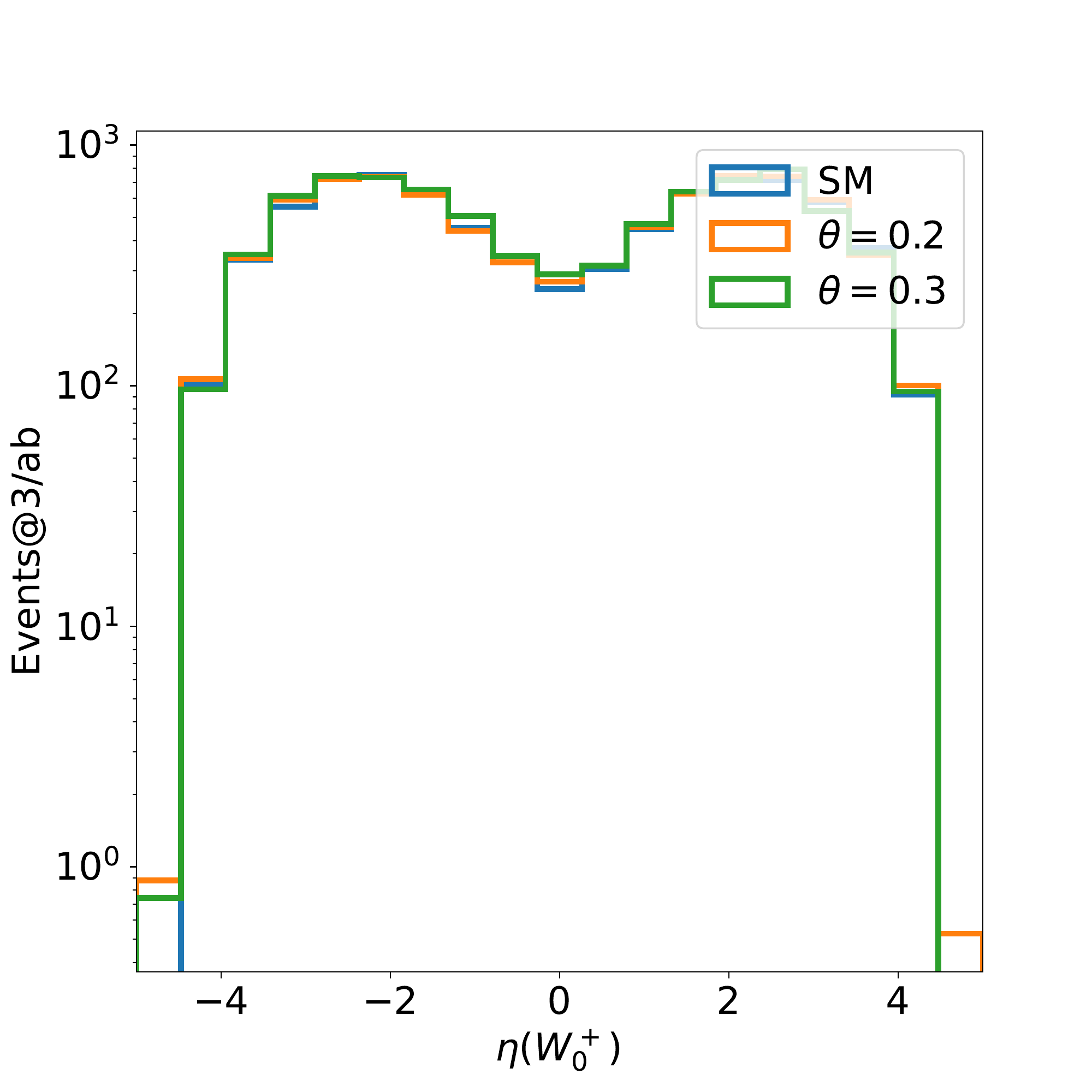}
	\includegraphics[width=0.32\textwidth]{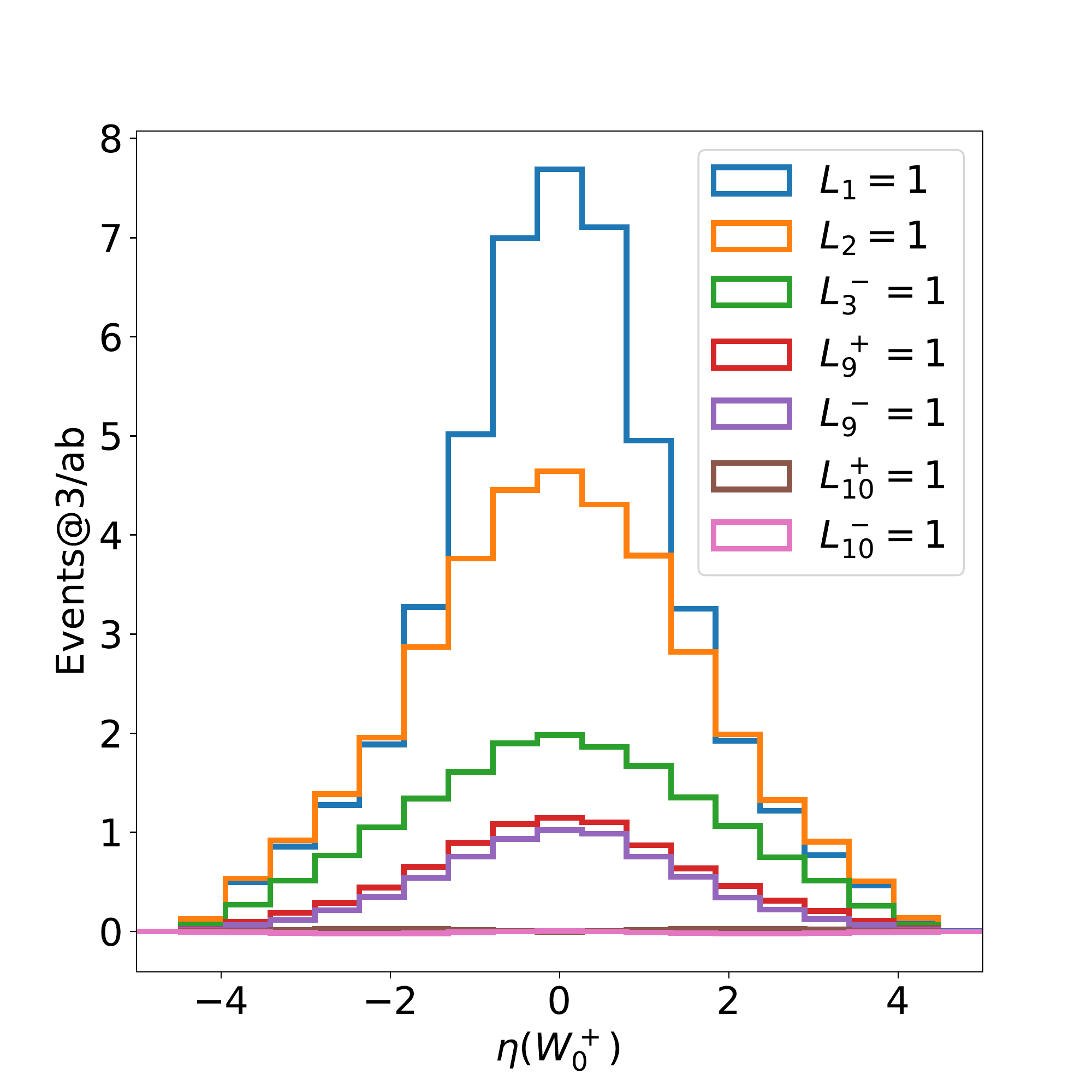}
	\includegraphics[width=0.32\textwidth]{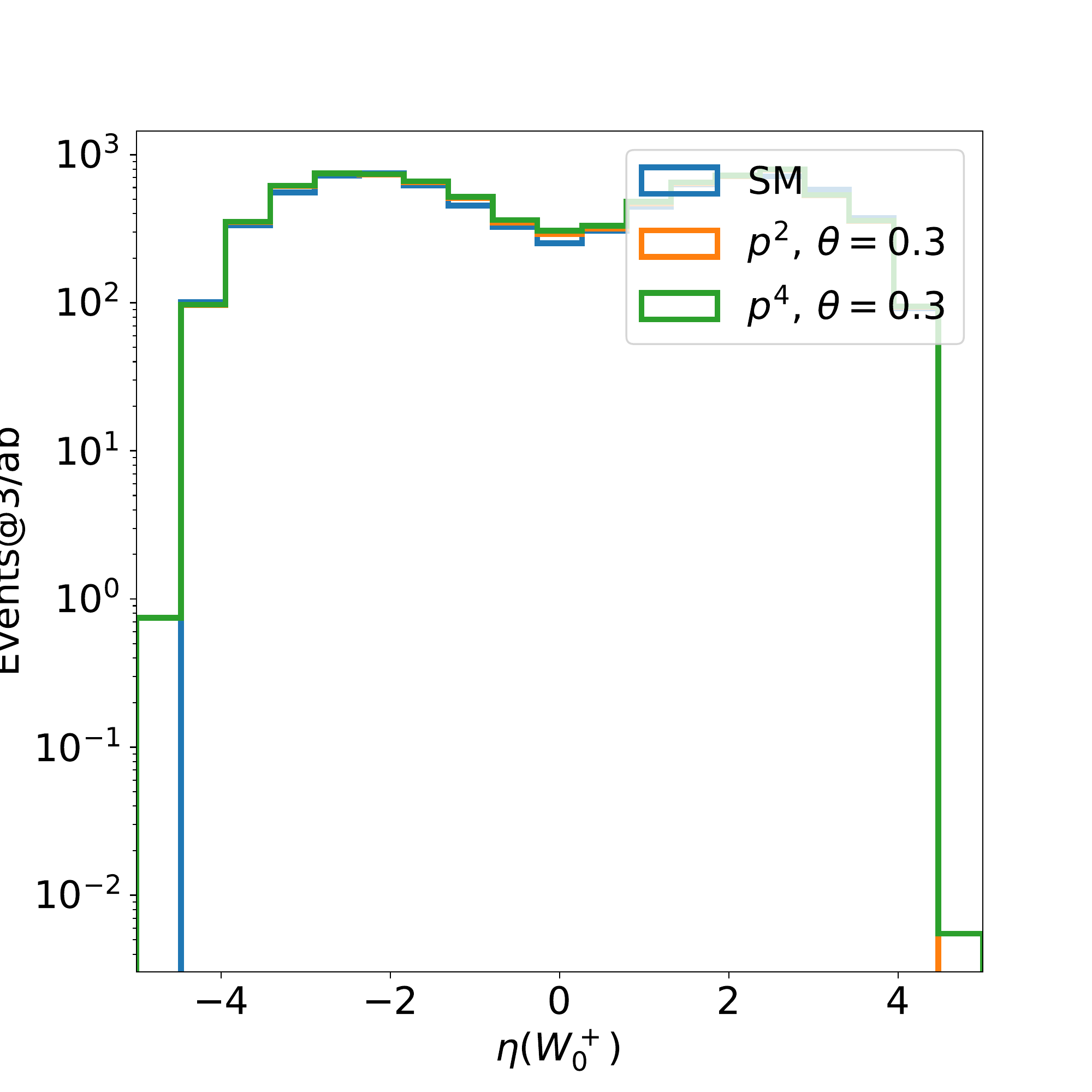}
	
	\caption{Longitudinal VBS.}
	\label{fig:vbs_pol}
\end{figure}

VBS $W^+W^-$ production $pp\to jj W^+W^-$
is depicted in diagram \fig{fig:diagrams} (right). 
We adopt the selection cuts shown in \tab{tab:cuts}.
Predictions for the distributions of the invariant mass of the $W^+W^-$  system $m(W^+W^-)$, the transverse momentum $p_T(W^+)$, and the pseudo-rapidity $\eta(W^+)$ of $W^+$ are shown in \fig{fig:vbs}.
The same format of \fig{fig:dihiggs} is used, with also similar observations about each NLO operator.
We also show predictions for the case where both final state $W^+$ and $W^-$ are polarized longitudinally. We used the polarized scattering implementation in \mg~\cite{BuarqueFranzosi:2019boy}. The corresponding distributions are shown in \fig{fig:vbs_pol}. It can indeed be noticed a larger new physics effect in the longitudinal components compared to the unpolarized scattering. This is expected since only the longitudinal component of $W^\pm$ can be identified as the NGB at high energies, according to the Equivalence Theorem.

\subsection{pNGB production and decay in non-minimal models}

We consider here two symmetry breaking cosets: $SU(4)/Sp(4)$ and $SU(5)/SO(5)$.
We include only the LO kinetic term and the WZW terms and neglect the SM fermion interactions.

\begin{figure}[htbp]
	\centering
	\includegraphics[width=0.45\textwidth]{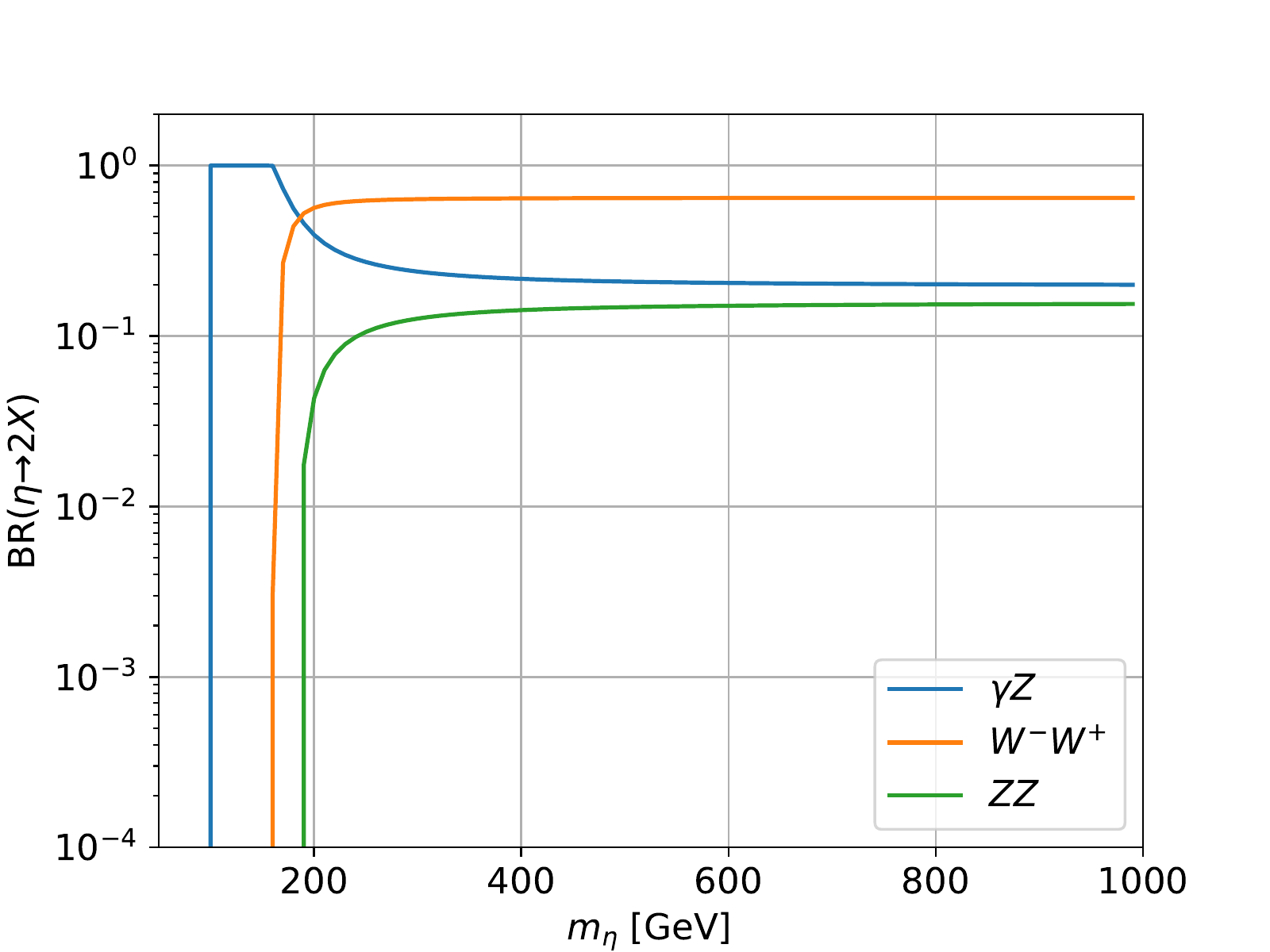}
	\caption{$\eta$ branching ratios in $SU(4)/Sp(4)$ model computed at tree level and including only 2-body decays via the WZW anomaly. }
	\label{fig:BRetaSU4Sp4}
\end{figure}

The $SU(4)/Sp(4)$ model contains 5 (p)NGBs. The Higgs bi-doublet {\bf (2,2)} and a singlet  $\eta$ {\bf (1,1)} under custodial
 $SU(2)_L\times SU(2)_R$ group.
The pair production of $\eta$ via VBF and associate production has been considered in several other publications (see e.g.~\cite{Arbey:2015exa}).
The branching ratios of $\eta$ are shown in \fig{fig:BRetaSU4Sp4}. We show only the 2-body decays computed with the analytic expressions provided in the UFO model. 
For $50\GeV \lesssim m_\eta\lesssim 90\GeV$ the offshell $Z^{*}\gamma$ decay channel dominates, and for $m_\eta\lesssim 50\GeV$ loop induced $b\bar{b}$ channel dominates.
See ~\cite{BuarqueFranzosi:2020baz} for more detail on the low $\eta$ mass case. 

\begin{figure}[htbp]
	\centering
	\includegraphics[width=0.32\textwidth]{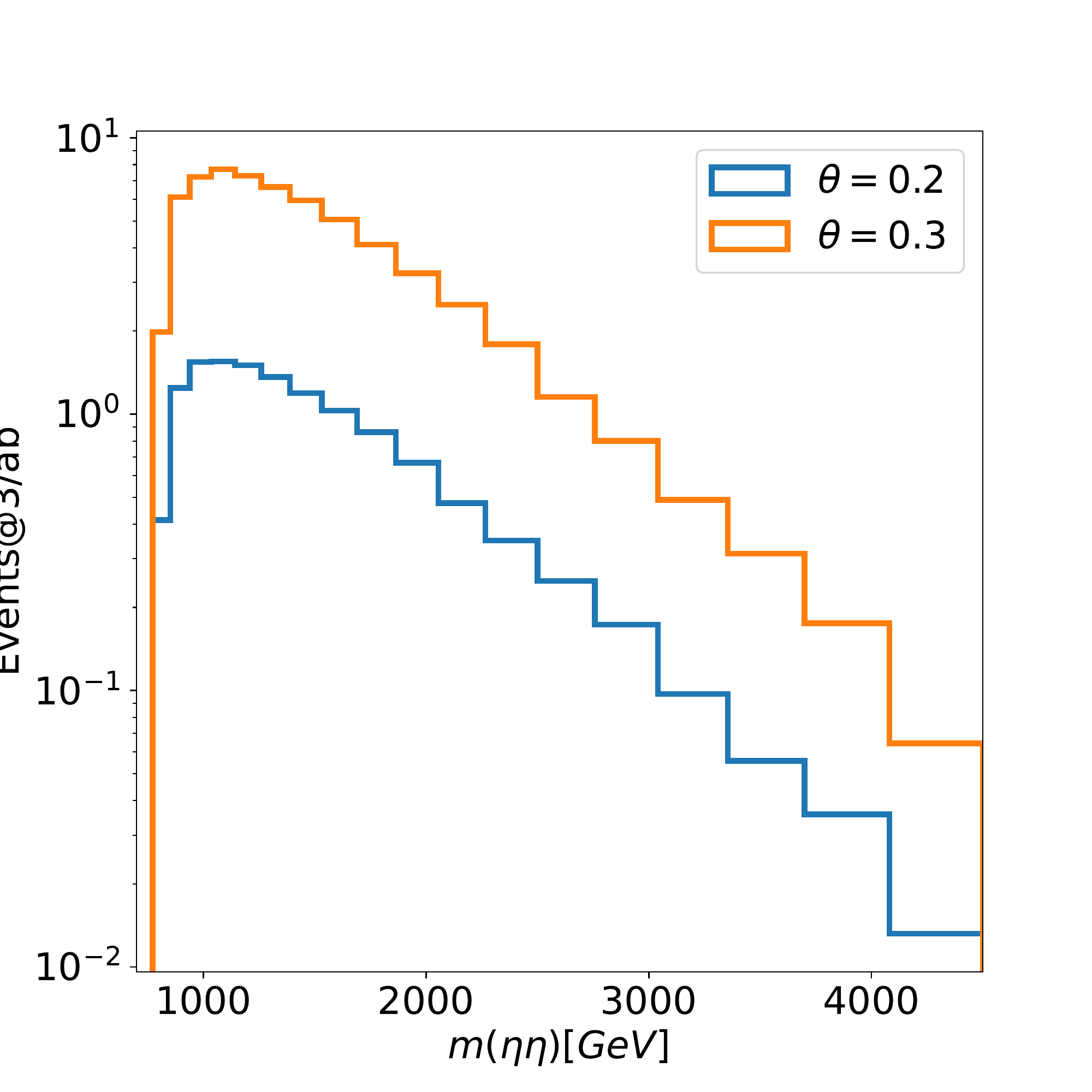}
	\includegraphics[width=0.32\textwidth]{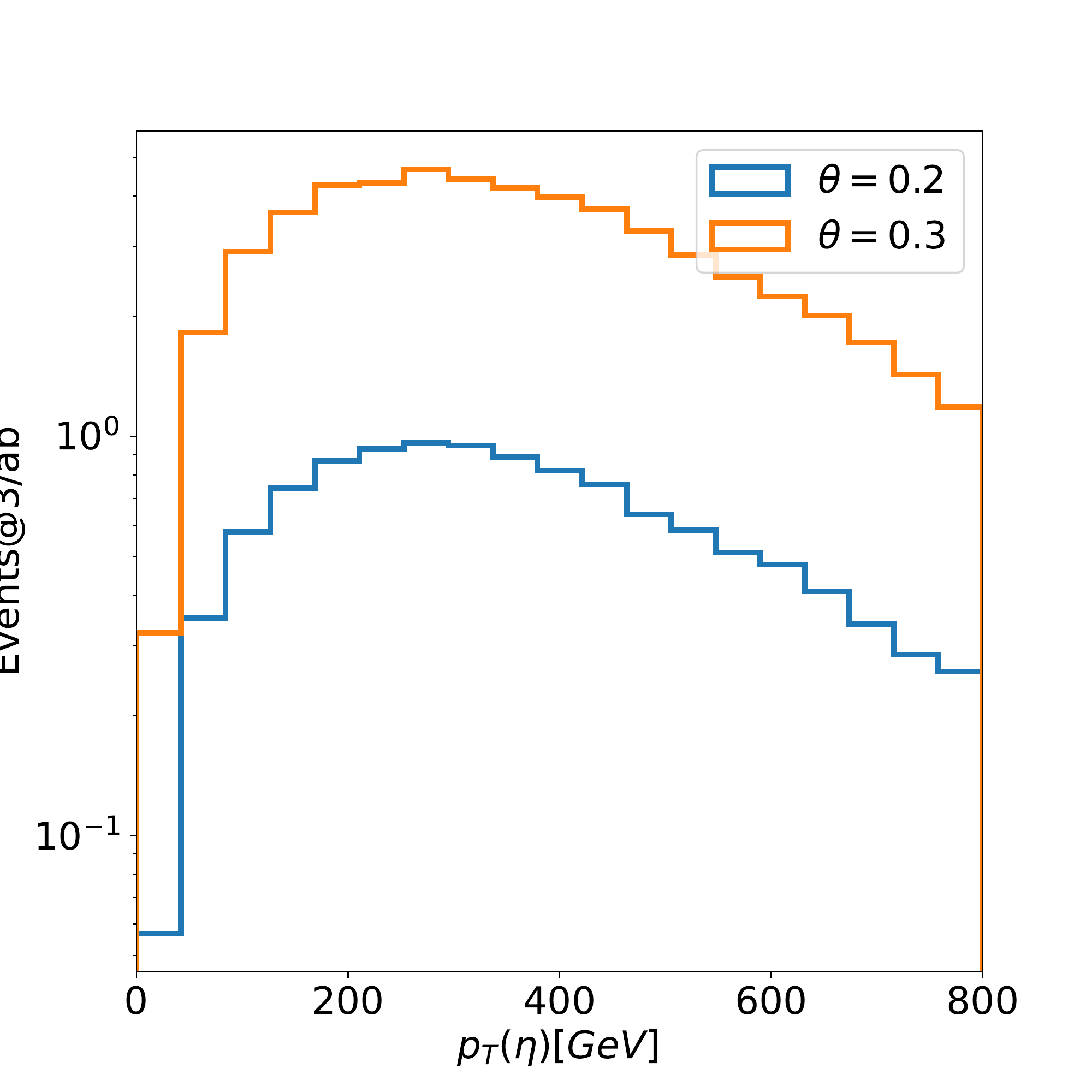}
	\includegraphics[width=0.32\textwidth]{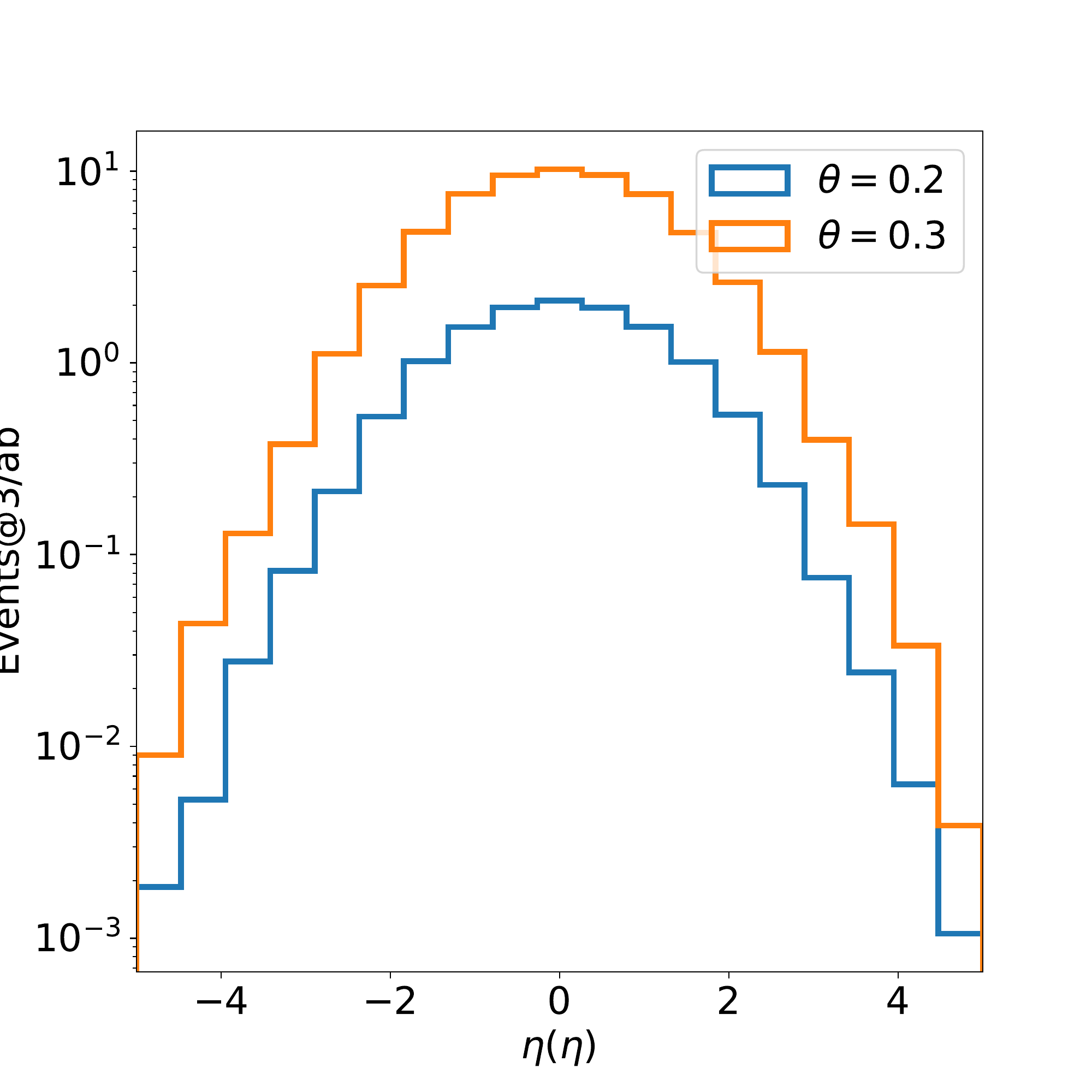}	
	\caption{$pp\to\eta\eta j j$ in $SU(4)/Sp(4)$. Left: the invariant mass of the $\eta\eta$ system. Middle: the transverse momentum. Right:  pseudo-rapidity of the hardest $\eta$.}
	\label{fig:etaetaSU4Sp4}
\end{figure}

Distributions for process $pp\to jj\eta\eta$ are shown in \fig{fig:etaetaSU4Sp4}. They are the invariant mass of the $\eta\eta$ system, the transverse momentum and pseudo-rapidity of the hardest $\eta$.
Event generation has been performed with the selection cuts shown in \tab{tab:cuts}.

The $SU(5)/SO(5)$ model contains ${\bf 14}\subset SO(5)$ (p)NGBs~\cite{Ferretti:2016upr,1808.10175}. ${\bf 14}$ decomposes into the Higgs bi-doublet, a bi-triplet and a singlet under the  custodial subgroup $SU(2)_L\times SU(2)_R$, ${\bf (2,2)+(3,3)+(1,1)}$.
We work in the custodial basis, further decomposing these multiplets into $SU(2)_V$, ${\bf (3,3)\sim 5+3+1}$, which we name $\eta_5$, $\eta_3$, $\eta_1$ respectively and the singlet ${\bf (1,1)}$ which we call $\eta$.
Typically there is mixing between these states, and the different charges of each $SU(2)_V$ multiplet can have different masses in general.

\begin{figure}[htbp]
	\centering
	\includegraphics[width=0.45\textwidth]{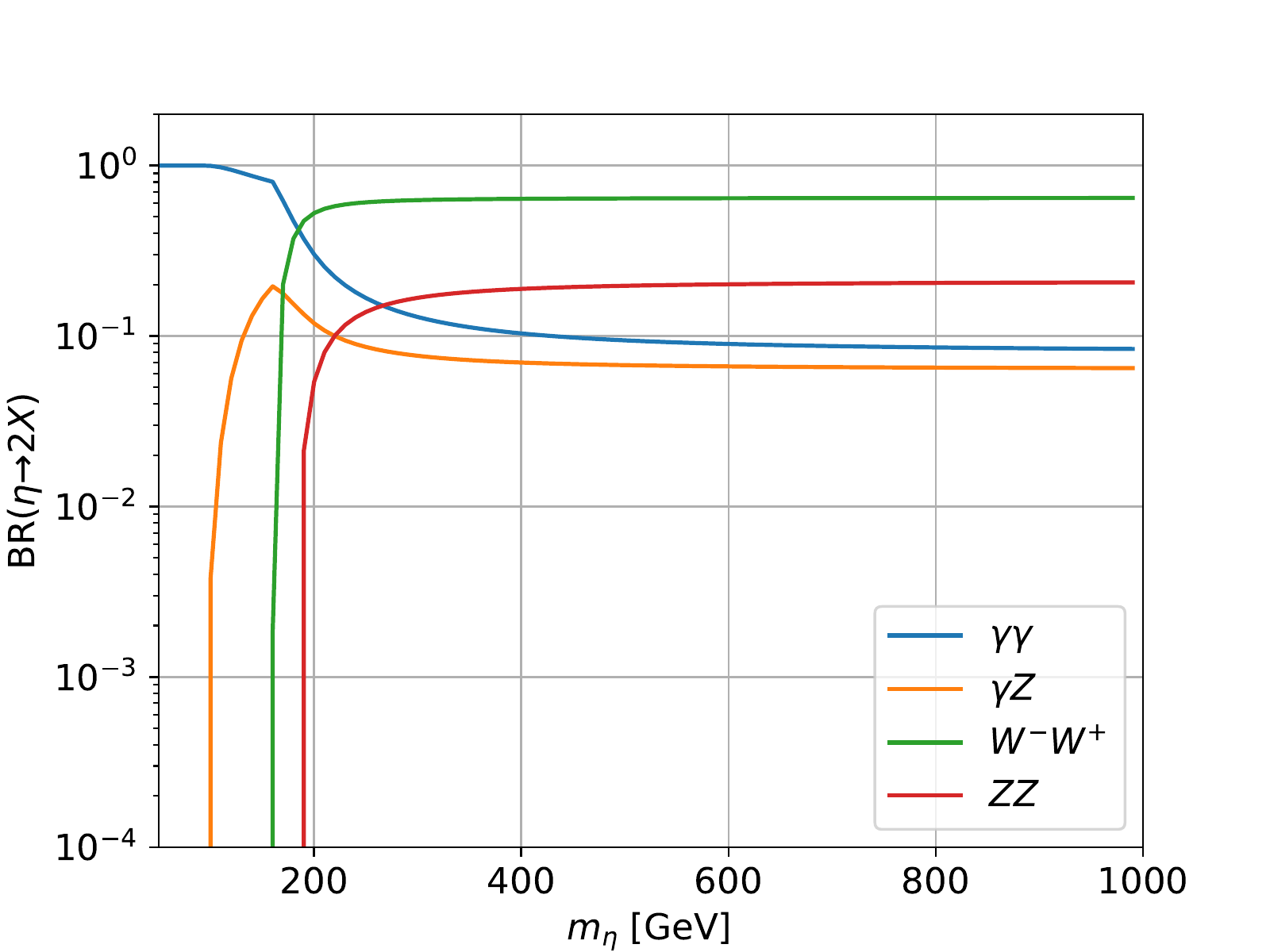}
	\includegraphics[width=0.45\textwidth]{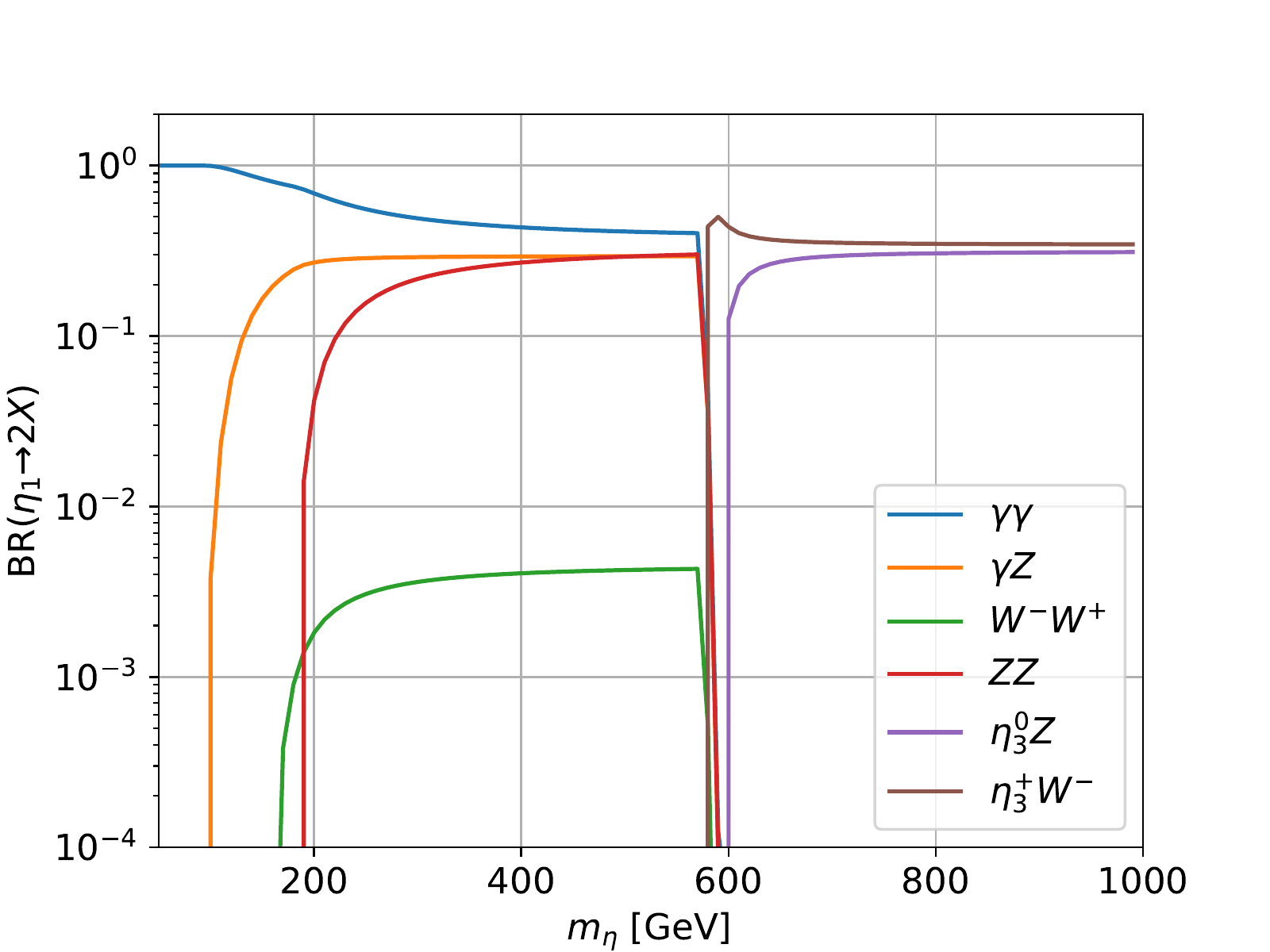}
	\includegraphics[width=0.45\textwidth]{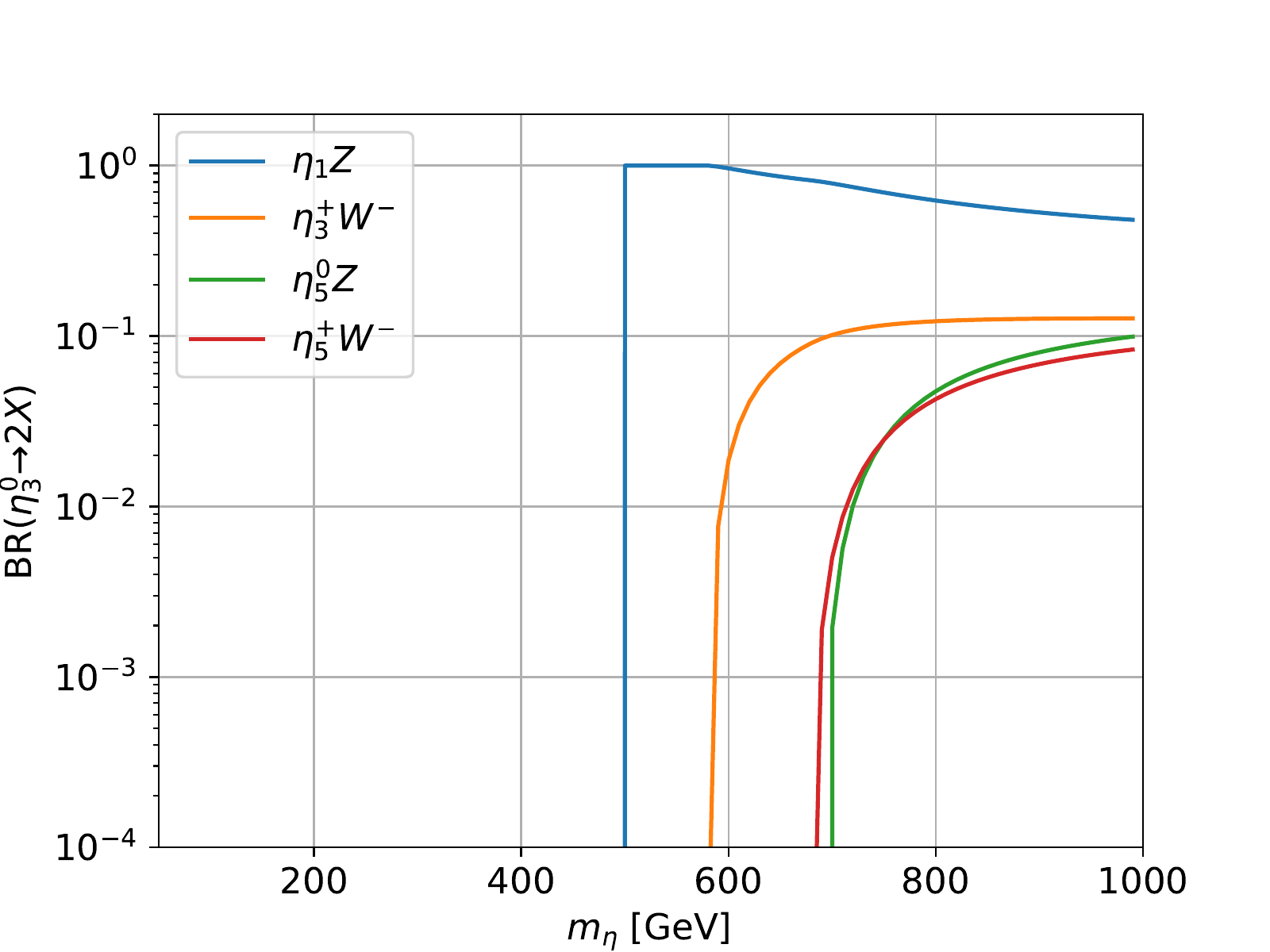}
	\includegraphics[width=0.45\textwidth]{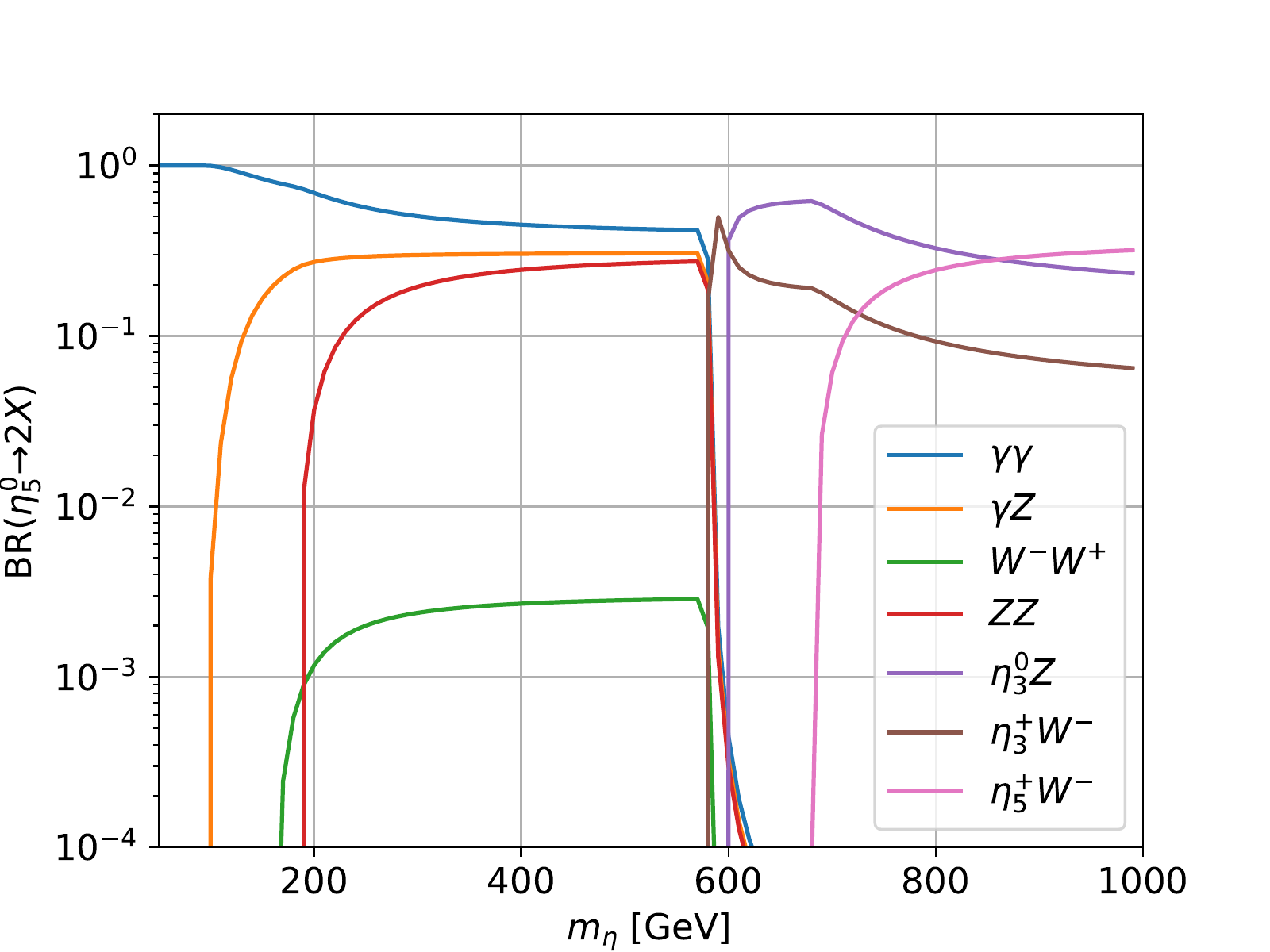}
	\caption{Two-body tree level branching ratios of neutral pNGBs in $SU(5)/SO(5)$. $\theta=0.3$, apart from one pNGB with scanned mass, the other pNGB masses are fixed to 
		$m_{\eta_5^{\pm\pm,\pm,0}}=600\GeV$, $m_{\eta_3^{\pm,0}}=500\GeV$, $m_{\eta_1}=400\GeV$, $m_{\eta}=300\GeV$.}
	\label{fig:BRneutral}
\end{figure}

\begin{figure}[htbp]
	\centering
	\includegraphics[width=0.45\textwidth]{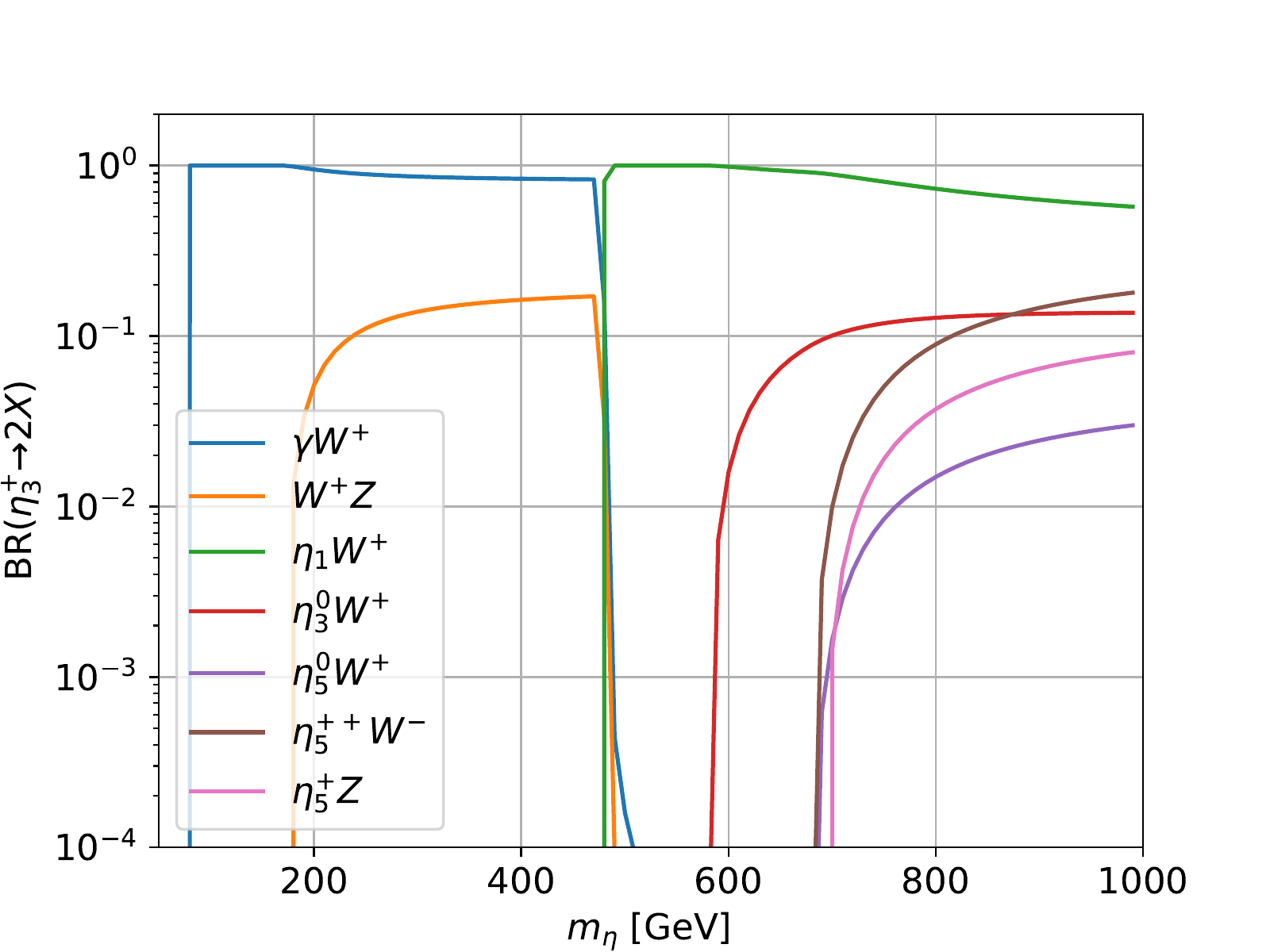}
	\includegraphics[width=0.45\textwidth]{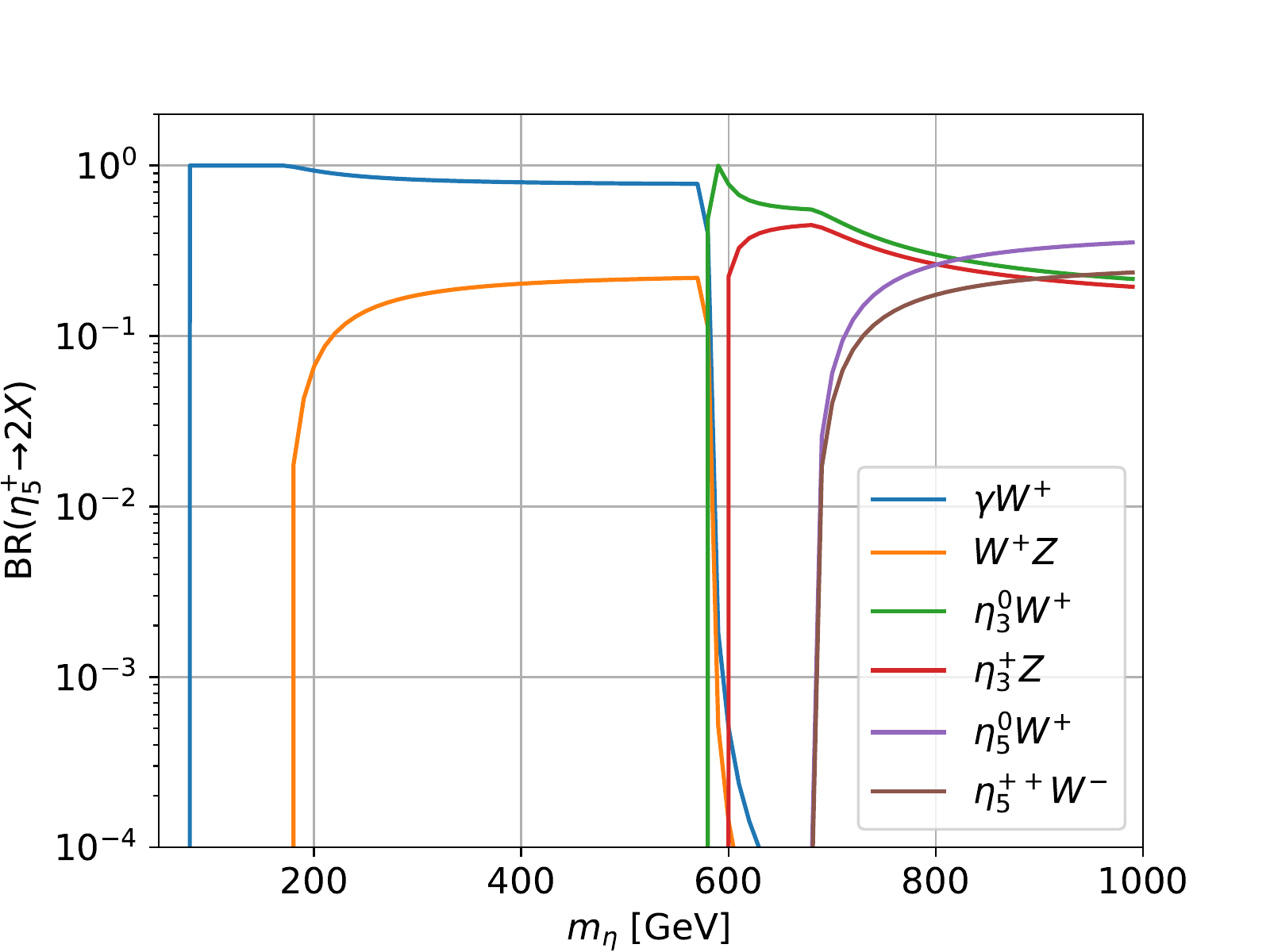}
	\caption{Two-body tree level branching ratios of charged pNGBs in $SU(5)/SO(5)$. $\theta=0.3$, apart from one pNGB with scanned mass, the other pNGB masses are fixed to 
		$m_{\eta_5^{\pm\pm,\pm,0}}=600\GeV$, $m_{\eta_3^{\pm,0}}=500\GeV$, $m_{\eta_1}=400\GeV$, $m_{\eta}=300\GeV$.}
	\label{fig:BRcharged}
\end{figure}

We further take as benchmark for the masses the following values:  
\begin{equation}
m_{\eta_5^{\pm\pm,\pm,0}}=600\GeV,\quad m_{\eta_3^{\pm,0}}=400\GeV,\quad m_{\eta_1}=300\GeV,\quad m_{\eta}=200\GeV\,.
\label{eq:pNGBmasses}
\end{equation}
The branching ratios into two body computed at tree level are shown in \fig{fig:BRneutral} for the neutral pNGBs and \fig{fig:BRcharged} for the singly charged ones. 
They are calculated for 2-body at LO only with the analytical expressions provided in the UFO model.
The  singlet of $SO(5)$ $\eta$ does not couple to other pNGBs and decay fully through the anomaly.
The other states usually decay into di-boson till a kinematic threshold into other pNGB opens up, and then tend to decay into them. 
Near kinematic threshold 3-body decays via off-shell propagating pNGBs can be important and are not shown in the figures. 

The $\eta_3^{\pm,0}$ state has no couplings via anomaly and decays only via the couplings to other pNGBs. If it is the lightest state, it decays to 3 body mediated by an offshell pNGBs.
As a example we take $m_{\eta^0}=300\GeV$ (keeping the other masses as in \eq{eq:pNGBmasses}) and get total width $\Gamma_3=0.32$ eV and the branching ratios  $\Gamma_{VVV}$ in \tab{tab:breta30}.

\begin{table}
\begin{tabular}{c|c c c c c}
	\hline
$VVV$ 				& $Z\gamma\gamma$ & $ZZ\gamma$ & $\gamma W^+W^-$ & $Z W^+W^-$ & $ZZZ$ \\
$\Gamma_{VVV}$ (\%)  & 74.17 & 14.25 & 10.64 &  0.73 & 0.21 \\
\hline
\end{tabular}
\caption{Branching ratios of $\eta_3^0$ for $m_{\eta^0}=300\GeV$, $\theta=0.3$ and the other masses as \eq{eq:pNGBmasses}. }
\label{tab:breta30}
\end{table}

As a example of GBS in this model we consider the doubly-charged scalar pair production via VBF $pp\to jj \eta^{++}\eta^{--}$.
The pair production of $\eta_5^{\pm\pm}$ is bounded by ATLAS searches, giving a lower bound on its mass $m_5\gtrsim 400\GeV$~\cite{ATLAS:2021jol}.
The Drell-Yan (DY) production and interpretation has been studied in \cite{Banerjee:2022xmu}.
For $m_5=400\GeV$ and $\theta=0.3$ the pair production production via VBF has total cross section $\sigma_{VBS}=0.135$ fb, to be compared with $\sigma_{DY}=5.38$ fb for Drell-Yan (DY) production. 
Event generation has been performed with the selection cuts shown in \tab{tab:cuts}.

\begin{figure}[htbp]
	\centering
	\includegraphics[width=0.45\textwidth]{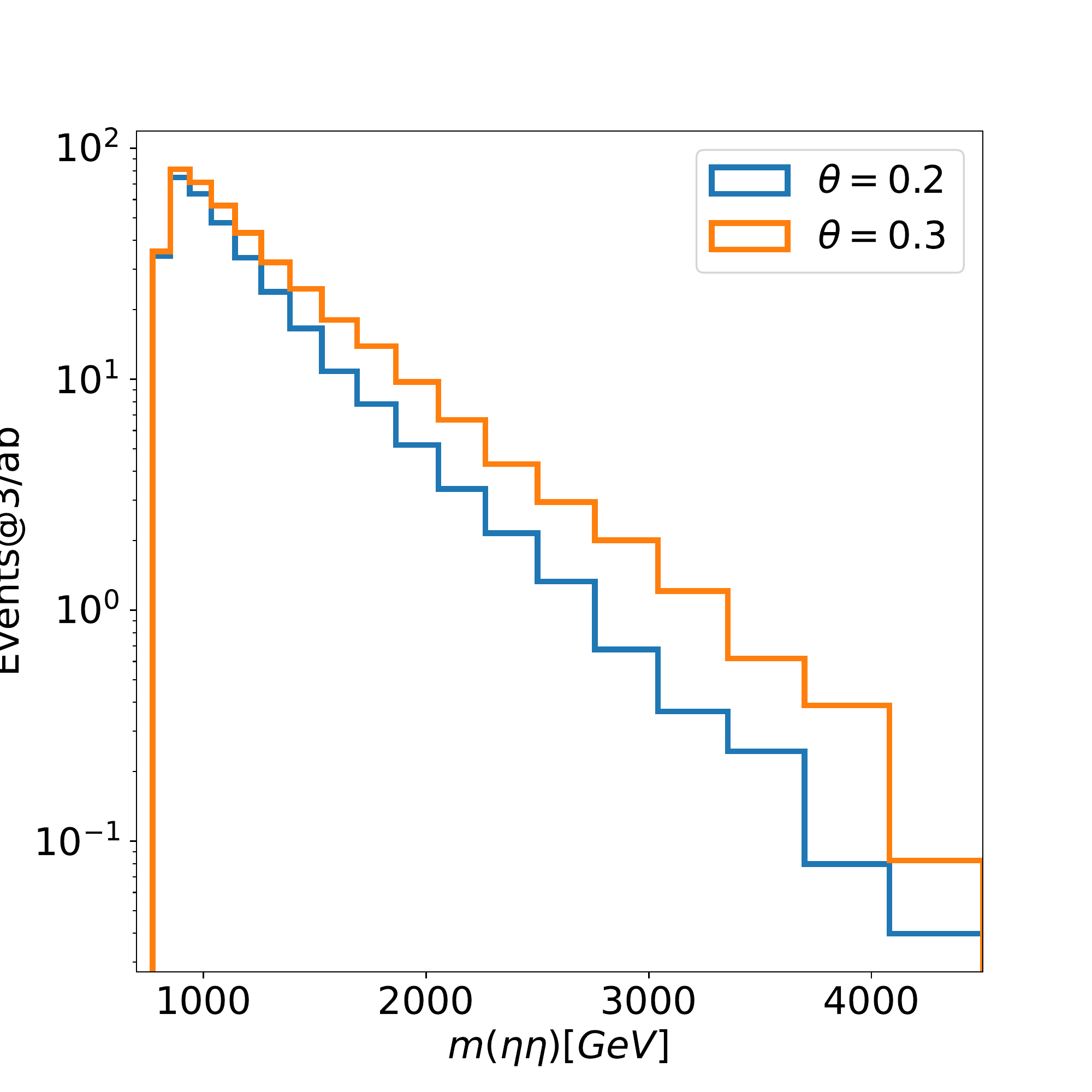}
	\caption{$pp\to\eta^{++}\eta^{--} j j$ in $SU(5)/SO(5)$. Invariant mass of the $\eta^{++}\eta^{--}$ system. }
	\label{fig:etaetaSU5SO5}
\end{figure}

 In \fig{fig:etaetaSU5SO5} we show the invariant mass of the $\eta^{++}\eta^{--}$ system.

\section{Conclusion and prospects}
\label{sec:conclusion}

We have presented a framework for the study of (p)NGBs in CH models, including VBS, di-Higgs via VBF, GBS in non-minimal models. 
We have implemented the model in FeynRules and output it in the UFO format. 
We then showed examples of simulations that can be further explored in more detailed phenomenological analysis.
We looked at the effect of $\mO(p^4)$ operators in VBS, in quartic couplings modifications and in di-Higgs production via VBF. 
We also computed the branching ratios and pair-production distributions for extra pNGBs in non-minimal models of CH for some particular benchmarks.

Besides the suitability for phenomenological studies, the tool here presented is intended as a first step towards the incorporation of loop corrections
of EW and chiral perturbation theory origin.

\appendix

\section{Explicit matrices for ``minimal cosets"}
\label{sec:group}

For $\SU(4)/Sp(4)$ we have $N=\sqrt{2}$ and $t=1/2$.
We use the matrices in Ref.\cite{Cacciapaglia:2014uja}. $\hat T_L^{1,2,3}=S^{1,2,3}$ $\hat T_R^{1,2,3}=S^{4,5,6}$. $X_h=X_4$.
\begin{equation}
\Omega=\exp(\ii N X_h \theta)=\cos\frac{\theta}{2} +\ii 2\sqrt{2} X_h \sin\frac{\theta}{2}\,.
\end{equation}

For $SU(4)\times SU(4)/SU(4)$ we get $N=2$, $t=1$.
Working with the $8\times 8$ matrices, we choose
\begin{equation}
\hat T_{L,R}^i = 
\left( \begin{array}{cc}
\tau_{L,R}^i & \\
& \tau_{L,R}^i 
\end{array} \right) 
\end{equation} 
with
\begin{equation}
\tau_L^i = 
\frac{1}{2\sqrt{2}}\left( \begin{array}{cc}
\sigma_i & \\
& 0 
\end{array} \right) \quad \text{and} \quad
\tau_R^i = 
\frac{1}{2\sqrt{2}}\left( \begin{array}{cc}
0 & \\
& \sigma_i 
\end{array} \right)
\end{equation} 

The Higgs generator can be chosen in one of 2  possible directions. We choose
\begin{equation}
X_h = 
\left( \begin{array}{cc}
\tau_H & \\
& -\tau_H 
\end{array} \right)
\end{equation} 
with
\begin{equation}
\tau_H = 
\frac{i}{4}\left( \begin{array}{cc}
0 & \uni  \\
-\uni &  
\end{array} \right)
\end{equation} 
So
\begin{equation}
\Omega = \exp\left[\ii 2\theta X_h\right] = \cos\frac{\theta}{2}+\ii 4 X_h \sin\frac{\theta}{2}
\end{equation}

For $SU(5)/SO(5)$ we have $N=2$, $t=1$.
The EW generators  can be chosen in the following way
\beq
T_L^i = 
\frac{1}{2\sqrt{2}}\left( \begin{array}{ccc}
	\sigma_i & & \\
	& \sigma_i & \\
	& & 
\end{array} \right) \quad \text{and} \quad
T_R^i = 
-\frac{1}{2\sqrt{2}}\left( \begin{array}{cc}
	\sigma_i^T
	\left( 
	\begin{array}{cc}
		\uni & \\
		& \uni  
	\end{array} \right) 
	& \\
	& 
\end{array} \right) 
\eeq 
The Higgs  is identified by the generator
\beq
X_{h} = 
\frac{i}{2\sqrt{2}}\left( \begin{array}{ccc}
	& & \binom{0}{1} \\
	& & \binom{-1}{0} \\
	(0,-1)  & (1,0) & 
\end{array} \right) 
\eeq
So
\begin{equation}
\Omega = \exp\left[\ii 2\theta X_h\right] = 1+ 4 X_h^2(\cos\theta-1) + \ii 2 X_h \sin\theta \,.
\end{equation}

\section{Explicit breaking terms and spurions}
\label{sec:spurions}

The gauging of the EW subgroup will explicitly break the global symmetry $G$ and the $j_\mu$ can be regarded as a spurionic field. 
Other interactions besides the gauge of $H'$ may also break  $G$ and can be included in the low energy chiral lagrangian via spurions. 
The only requirement is that the local SM gauge group be unbroken, and preferably that the whole custodial symmetry be respected.
Following \cite{Alanne:2018wtp} we concentrate here on spurions transforming in the fundamental (F), two-index symmetric (S) or anti-symmetric (A) and the adjoint (Adj) representations.
The best example of such terms is given by hyperfermion masses $\bar{\psi}\mathcal{M}\psi$, which transforms in general as $\Sigma$ (\label{eq:sigma}),
\begin{equation}
\mathcal{M}\to g \mathcal{M} g^T,\quad \chi =\xi^\dagger \mathcal{M} \xi^*\ez \to h\chi h^\dagger
\end{equation}
Notice that with our way to write the $SU(N)^2/SU(N)$ case, we can keep the same structure of a simple group, without having to specify $g_L$ or $g_R$.
The other spurions representations transform as
\begin{eqnarray}
\Xi_F\to g\Xi_F,\quad \Xi_{S/A}\to g\Xi_{S/A}g^T,\quad \Xi_{Adj}\to g\Xi_{Adj}g^\dagger.
\end{eqnarray}
The A/S and Adj representation can be put together in the object
\begin{equation}
\chi = \begin{cases} \xi^\dagger \Xi_{A/S}\xi^*\ez  \\
\xi^\dagger \Xi_{Adj}\xi \end{cases} \to h\chi h^\dagger
\end{equation}
For the symmetric and anti-symmetric cases, since we always want to add the hermitian conjugate of the terms we can define a hermitian and an anti-hermitian combination
\begin{equation}
\chi_\pm = \frac{1}{2}(\chi\pm \chi^\dagger)\,.
\end{equation}
For the adjoint $\chi_+=\chi$ and $\chi_-=0$.
With those objects we can construct different objects transforming homogeneously under $H$ ($O\to h O h^\dagger$) that can be used to construct invariants for the Lagrangian, such as
\begin{eqnarray}
\langle \chi_+\rangle,\, \langle \chi_\pm^2\rangle,\, \langle \chi_\pm\rangle^2, \,\langle x_\mu \chi_+ \rangle,\\
\langle \chi_+\epsilon_0\chi_+^T\epsilon_0\rangle,\,\langle \chi_-\epsilon_0\chi_-^T\epsilon_0\rangle,
\end{eqnarray}

\section*{Acknowledgements}
We thank Gabriele Ferretti and Avik Banerjee for reading of the manuscript and for useful comments.
We thank Ilaria Brivio for relevant discussions about the kinetic normalization of the gauge bosons.
Computational resources have been provided by the supercomputing facilities of the Université catholique de Louvain (CISM/UCL) and the Consortium des Équipements de Calcul Intensif en Fédération Wallonie Bruxelles (CÉCI) funded by the Fond de la Recherche Scientifique de Belgique (F.R.S.-FNRS) under convention 2.5020.11 and by the Walloon Region.

\bibliography{CHLag.bib}

\end{document}